# Sensors as Information Transducers

J. David Zook, Dresden Associates,
2540 Dresden Lane, Golden Valley, MN 55422 (763-529-5185)
Norbert Schroeder, Intechno Consulting,
Steinenbachgaesslein 49 CH-4051, Basel, Switzerland (++41) 61 281 18 30

**CONTENTS**

## 1. OVERVIEW

This chapter reviews the mechanisms by which sensors gather information from the physical world and transform it into the electronic signals that are used in today's information and control systems. It also introduces a new methodology for describing sensing mechanisms based on the process of information flow and applies it to the broad spectrum of the sensors, instruments and data input devices in current use. Specific examples are used to illustrate how information is transformed as it flows through a sensing subsystem in well-defined steps to become an electronic signal. Each step is associated with an elemental transduction process (ETP) and a given sensing mechanism that can be described in terms of the sequence of ETPs in the information path.

We identify four distinct ETPs: energy conversion (C), energy dispersion (D), energy modulation (M) and modulation of a material property (P). Describing a particular sensor in terms of its ETP sequence provides an understanding of its sensing mechanism and of its strengths and limitations as a selective gatherer of information. We posit that these four mechanisms form a complete set for describing information transduction in sensing systems. This hypothesis is tested and illustrated by applying it to about 100 kinds of commercially available sensors along with an estimate of the 2003 global market for each kind. The methodology also applies to other input devices, measurement instruments and sensing systems that gather information for today's information and control systems but are not part of the sensor literature or marketplace.

**SENSORS AND INFORMATION**

This section considers relationships between sensors and information. The purpose is to establish clear terminology for analyzing and classifying sensing mechanisms. The framework is established by considering sensing processes as forms of information processing that include the transduction of information. A hurdle to be overcome is the definition of "sensor." Where does the sensor start and where does it end in the chain of information processing? From a practical point of view, the term "sensor" is defined in terms of the sensor products available or potentially available in the sensor marketplace. However, a sensor product may include signal processing that is not part of the sensing mechanism and it may not include some elements essential to the sensing mechanism. We use "sensing system" as a broad term referring to a complete system that gathers information about events or qualities of the physical world and transforms it into the electronic signals that are used in conventional signal processing systems. These include sensors but also other information input devices or systems that are not part of the sensor marketplace. The concepts of "information transduction" and "information transducers" emphasize the transformation of information as it flows from the outside world to become an electrical signal.

### 1.1 Information.

Each person has a strong sense of what information is and of its importance in his or her life. Information is something that can be communicated, that can be stored and that can take many forms - the DNA in each living cell is one form of stored information. It is important to define information as it applies to sensing systems. The present discussion addresses the information in technology-based systems, as opposed to nature-based systems. We take the liberty of defining "information technology (IT) systems" very broadly to include all technology-based data processing systems such as computers, telecommunications and control systems that deal with analog or digital data. Information is the quintessential "stuff" of IT, the data that IT systems work with and (we assert) it is all provided by sensing systems. Sensor information must have two attributes to be useful: (1) it must describe a measurable attribute of the physical world external to the IT system, and (2) it must be distinguishable from noise (random signals). Thus we can define sensor information as the useful signal (describing an attribute of a measurand) that is distinguishable from a noise background.

Useful signals come in many forms: optical, acoustic, magnetic and chemical as well as electrical [1-3]. From a practical standpoint, however, IT systems only make use of electrical signals, whether digital or analog. Even though they use, store and retrieve information (on magnetic or optical discs, for example) the stored information is transformed to electrical signals for processing, analyzing and coding using the techniques of electronic signal processing. Information theory [4-9] defines the amount of information in an arbitrary set of data as the *minimum* number of bits that the data can be reduced to without loss of meaning of the message. As part of an information network a sensor is a Gaussian broadcast channel [7]. From the sensor point of view it is the information channel capacity (bit rate) that is an important figure of merit. Most sensors have a continuous, or analog, output that covers a certain dynamic range, which amounts to a certain number of bits of precision. The dynamic range and bandwidth (or response time) together determine the bit rate capability of a sensor. For example, a temperature sensor generates a continuous signal that can change in a time determined by the bandwidth of its readout system. A "smart" temperature sensor [1-3] might send a signal only when the temperature changes by a predetermined increment and thus the signal can be compressed into many fewer bits. Most sensor products do not include such data compression, so that, most of the time, their output includes many bits that are part of the information flow (above the noise level) but are not part of a minimum message. Our definition of information reflects the common practice of quantifying information in terms of information capacity rather than meaningful information. For example, a computer hard drive may contain 15 GB of stored data, but a much smaller amount of meaningful information.

Since signals can come in a number of forms and can be stored in a number of ways it is clear that "information" is a very broad concept. This chapter considers information as a sensor-based phenomenon. That is, information must enter an IT system through a sensor or sensor-like input device. By understanding sensing mechanisms more deeply and broadly, it may be possible to gain a better understanding of the nature of information.

### 1.2 Terminology related to information transducers.

An IT system (an information processing or control system) can be viewed as a chain consisting of three parts: an input part, a modification or processing part and an output part [3]. Inherent in this viewpoint is the concept of information flow between the system and the world outside the system. At the core of the system is the information processor. Generally information processing implies electronic signals or data. The input part then consists of transducers that gather and transform information from the physical world with the last step of the transformation resulting in an electrical signal. Since this paper addresses sensor mechanisms it is convenient to consider any interface electronics associated with the input process to be part of the signal processing system. Examples of devices that might be in each of the three parts are given in Table 1.

**Table 1. The three parts of a (broadly defined) IT system**

| Input Transducers | Information Processors | Output Transducers |
|---|---|---|
| Touch-based: | Computers | Displays: CRTs, LCDs, etc |
| keyboards | Amplifiers, filters, limiters | Printers |
| computer mice | Modulators, mixers | Write heads for storing data |
| Read heads for | Controllers (PIDs, etc.) | Loudspeakers, headphones |
| stored data | Counters | Control valves |
| Cameras, scanners | Multiplexers | Pneumatic controls |
| Data instruments | A/D converters | Hydraulic controls |
| Sensors | Fourier transforms (FFTs) | Motors, pumps, fans |

Some observations can be made: (1) Information comes from outside the system. It is meaningful only when it represents some attribute of the physical world outside the information system. (An exception is information that originates within the system, e.g., for self-diagnosis). (2) Sensing is an information transduction process; signals from outside the system are transformed into electrical signals that can be processed by a data or signal processing system. (3) The sensing process is (non-electronic) information processing that takes place when some attribute of the physical world is converted to an electronic signal. (4) To understand sensing mechanisms it is important to consider the entire sensing or transduction process that may include processes occurring outside the "sensor" or the information system. (5) Sensing cannot add information but can only reduce the amount of information available (by adding noise or distortion or by limiting bandwidth).

Most books on sensors [10-26] fail to make a connection between sensors and information. In order to make that connection we have found it useful to make a departure from the conventional definition of transducer. Gershenfeld expresses the common understanding related to sensors as transducers:

> "*Sensors* are used to convert a parameter of interest to a form that is more convenient to use. The most familiar sensors provide an electrical output, but many other kinds of mappings are possible. … Sensors are sometimes called *transducers*, although transduction is more correctly used to refer to bidirectional mechanisms such as piezoelectricity that can convert between pressure and charge." ([9], p.231)

The term *information transducer* is an inclusive term that covers any mapping, any transformation of information from one form to another. As we point out with many examples, it is a much broader term than sensor and labels a concept essential to discussing the sensing process and analyzing sensing mechanisms. Contrary to the above description of *transducer*, information transducers generally transform information in only one direction, i.e., have a specified input and output.

With this definition, sensors are simply a subset of the larger set of information transducers. There are three categories of information transducers – input transducers, information processors and output transducers (See Table 1), depending on where the transducer is in the information chain. Input transducers include sensors but also include other types of information transducer that are not usually considered sensors – in the marketplace or in the literature. "Sensors" refers to input transducers that measure some attribute of the physical world, as we discuss in more detail below. "Touch-based transducers" have a different purpose and are usually part of a man-machine information loop that uses human touch for data entry. In this category, keyboards, computer mice and touch-screens are not complete input devices by themselves but are part of a feedback loop that includes an output device such as a CRT or LCD screen. Another category of input transducer has a still different purpose – reading optically or magnetically stored data. Read heads for stored data make use of sensor mechanisms but are not necessarily part of the sensor marketplace or sensor literature. We will see that all information-input transducers make use of the same four information transduction mechanisms.

Since information can come in many forms (electrical, optical, chemical, etc.) and types (binary, digital or analog), information transducers can come in many combinations. The present discussion centers on IT input transducers – devices that generate electrical output signals compatible with today's information processing and control systems. The focus is on current sensor products. The concepts of information and information transduction are much broader. They could include purely mechanical, pneumatic or hydraulic control systems, for example. Perhaps future sensors and information transducers will be even more diverse, making use of optical data processing, quantum computation or even neurotransmitter receptors.

### 1.3 Sensors and sensing subsystems.

The definition of *sensor* requires discussion and clarification. Historically it has included devices with a variety of labels. Jones notes that:

> "Measurement technology embraces a wide vocabulary with poorly defined meanings. A device that can perform a measurement conversion may be called a sensor, sensing element, transducer, transmitter, converter, detector, cell, gauge, pick-up, probe, transponder or X-meter where X is any quantity to be measured. Generally, though, the terms 'sensor' or 'transducer' are now used to describe a device for the measurement of a physical or chemical quantity by electrical means." ([12], p.3)

The trend has been toward the preferred use of the word "sensor," leaving the use of "transducer" for energy transducers only, consistent with Geshenfeld's definition above. Neubert gives a definition of "sensor" that seems hard to improve upon:

"[Sensors] are devices which, for the purposes of measurement, turn physical input quantities into electrical output signals, their output-input and output-time relationship being predictable to a known degree of accuracy at specified environmental conditions." ([10], p.1)

Interestingly, Neubert (in 1975) was actually defining "instrument transducers" rather than "sensors," so the above quote is a deliberate misquote to illustrate how the terminology has changed. Both the Neubert and Jones definitions of sensors emphasize the measurement aspect of sensors and indeed, for sensors that have the purpose of measurement, it is appropriate to call the input variable a "measurand." However, some sensors have other purposes, as discussed below.

One reason for the increased use of the simple term "sensor" (as opposed to "transducer") is that the numbers and kinds of sensors have grown as information technology has grown. Today's technology has enabled the use of an increased variety of sensors responding to more kinds of physical, chemical and environmental variables. For example, the kinds of sensors used on automobiles [18] has expanded very rapidly, in response to the need for safety (seat belt, door lock sensors) and improved performance and fuel economy (mass air flow, oxygen sensors).

Fraden gives a broader definition of sensor:

"A sensor is a device that receives a signal or stimulus and responds with an electric signal." ([27], p.2)

He goes on to define "stimulus" as "the quantity, property, or condition that is sensed and converted into electrical signal." This broader definition includes devices whose purpose is not measurement but rather to provide a binary output when a specific "event" occurs. Examples are glass-breakage detectors and fire detectors. Their purpose is not measurement of an analog variable, but rather is the detection of an event and the initiation of an alarm signal (or the inflation of an airbag) if an event is detected. Such devices are considered as sensor products in the sensor marketplace [28] and are thus included in our definition of sensors.

In other respects, Fraden's definition is too broad. It includes man-machine interface devices such as computer keyboards that are not considered sensors or part of the sensor market. On the other hand, these devices are clearly in the broader category of information input transducers. In Table 1 these are identified as "touch-based transducers." Like event sensors their purpose is not measurement. Since their purpose is not quantitative measurement the term "measurand" does not apply. As Fraden points out, the term "measurand" over-emphasizes the quantitative aspect of sensing. On the other hand, his use of the word "stimulus" does not seem appropriate for some kinds of sensors (such as the "Type M" sensors described below). From the viewpoint of information transduction the terms "input parameter," "input variable" or "input signal" are preferable.

This chapter uses a pragmatic definition of "sensor" based on the sensor marketplace and the sensor literature: if it is called a sensor, it is a sensor. In the sensor marketplace "sensor" refers to the package that is sold as a sensor product. Thus a "sensor" may be an unpackaged sensor chip or it could be a product including additional electronic functions such as signal conditioning electronics, analog-to-digital converters, circuits for linearizing the response function and even microprocessors in the same package, making them "smart sensors." A product that has its own display but no provision for output to a control system or display is considered an "instrument," not part of the sensor marketplace. Touch-based data input devices and clocks make use of the same types of information transducers as sensors but are not considered sensors. Discussions of sensing mechanisms, on the other hand, need not include signal processing even if it is part of the sensor product.

Generally, the marketplace definition is more restrictive than the literature descriptions: a product is only marketed as a sensor if it can claim to provide a robust electrical output without human supervision or intervention. This is clearly a requirement for sensors in autonomous control systems. These applications generally require sensors that are unattended; they must gather information without human intervention. Sensors for laboratory instruments or medical diagnostic tests, on the other hand, provide information to be displayed and/or stored for human interpretation. Chemical sensors and biosensors, for example, may require human intervention for calibration before and after a measurement and may be disposable.

Another complication around the definition of "sensor" is that additional elements may be required for the sensing process (the gathering of information) but they may not be included in the marketed package. Optically based sensors provide some excellent examples. Image sensors require a lens to direct the light to each pixel of the sensor array, but the lens may or may not be part of the sensor product. The source of light is also essential to the sensing process in imaging sensors and in optical scanners but is usually not considered part of the sensor. An exception here is the ring laser gyro in which laser is an intimate part of the sensing mechanism. Thus there can be apparent conflicts between the pragmatic market definition of "sensor" and the basic mechanisms involved in the sensing process. For clarity, the term "sensing process subsystem" or "sensing subsystem" will be used to refer to complete set of elements, structures or devices involved in gathering information about the physical world and transforming it into an electrical output signal. (See Figure 1.)

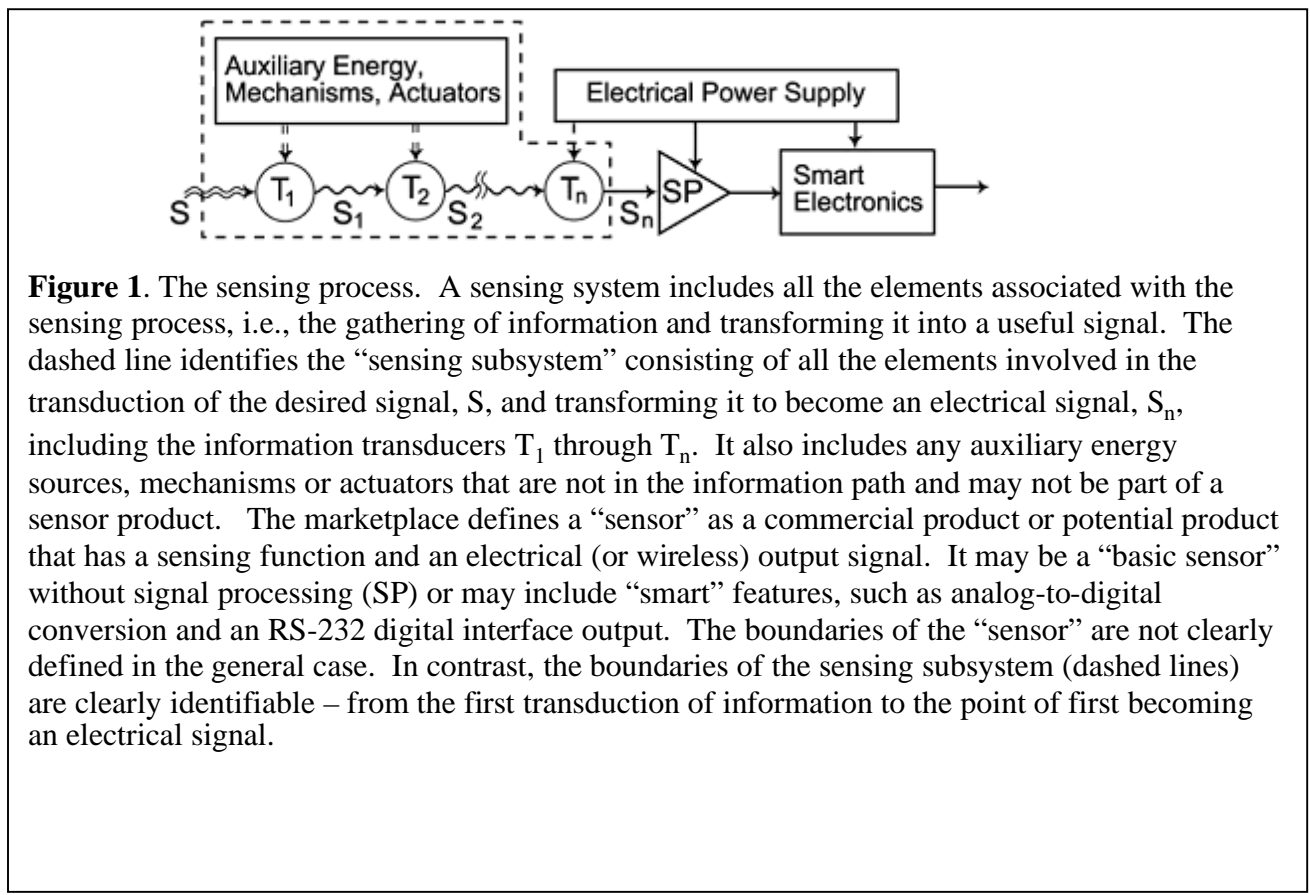

**Figure 1**. The sensing process. A sensing system includes all the elements associated with the sensing process, i.e., the gathering of information and transforming it into a useful signal. The dashed line identifies the "sensing subsystem" consisting of all the elements involved in the transduction of the desired signal, S, and transforming it to become an electrical signal, $S_n$, including the information transducers $T_1$ through $T_n$. It also includes any auxiliary energy sources, mechanisms or actuators that are not in the information path and may not be part of a sensor product. The marketplace defines a "sensor" as a commercial product or potential product that has a sensing function and an electrical (or wireless) output signal. It may be a "basic sensor" without signal processing (SP) or may include "smart" features, such as analog-to-digital conversion and an RS-232 digital interface output. The boundaries of the "sensor" are not clearly defined in the general case. In contrast, the boundaries of the sensing subsystem (dashed lines) are clearly identifiable – from the first transduction of information to the point of first becoming an electrical signal.

An unexpected feature of the "sensing subsystem" concept is that it can include the object being sensed. This is necessarily the case when stored information is transformed into an electrical signal. An example is an optical barcode scanner. The barcode is part of the physical world outside the scanner but the reflection of laser light from it is the key transduction process in the sensing subsystem. The laser is an example of an auxiliary energy source shown in Figure 1 and contains no information about the input signal (the barcode). The first transduction process is the reflection of light from the barcode. Thus the signal information S is the barcode information that was stored when the barcode was printed. It makes sense that the barcode is part of the sensing subsystem but of course *not* part of the sensor. Other examples are given below (for example, see Box 3.)

    **2.4 Information flow.** The concept of directed information flow that is implicit in Figure 1 is inherent to communication, signal processing and information theory [7]. Each stage of signal processing is described in terms of an input port and an output port (and a transfer function relating them) associated with the direction of information flow. The analysis of the information flow *within* a sensor or a sensing subsystem is straightforward but seems not to have previously been discussed in the literature. Jones comes closest with his concept of the "primary-sensing element" [12]. He uses this term to refer to the first interface between the sensor and the outside world, the "front end" of the sensing system, the $T_1$ transducer in Figure 1. He gives some examples: diaphragms, bellows and Bourdon tubes for measuring pressure, bimetallic strips for temperature, floats and weirs for level, turbines, Venturi tubes, orifice plates and bluff bodies for flow, pitot tubes for air velocity. These structures are not sensors by themselves, even though the "performance [of the sensor] is still highly dependant on that of the front-end, primary-sensing element."

    The present viewpoint is that the information flow in a sensing subsystem can be broken down into sequential elementary transduction processes (ETPs). Jones' primary-sensing elements are examples of ETPs and a complete "sensing subsystem" consists of the sequence of ETPs that result in an electrical signal as illustrated in Figure 1. The ETP concept is simple but not obvious because ETPs can be implemented in simple structures such as diaphragms, bimetal cantilevers and electrical switches that have not been recognized as part of the information flow. The diversity of ETPs is much greater than the number of measurands since a wide variety of structures and devices can act as information transducers. Perhaps this is not surprising – even a leaf fluttering in the breeze is transducing information about that airflow, and as such is an example of a (non-IT) ETP.

## 2. PREVIOUS CLASSIFICATIONS OF SENSORS

    Sensors can be classified in a variety of ways, for example by measurand, by technology or by sensing mechanism. These classifications are not necessarily related. For example, it is possible to have very different mechanisms for the same measurand – a speedometer or a Doppler radar can measure velocity, or it can be derived as the rate of change in distance with respect to time, or from an accelerometer measurement by integrating with respect to time. As the content of this encyclopedia illustrates, there are a wide variety of sensors, measurands and ways to classify them. General references on sensors [3, 9-27] classify them by measurand but then each seems to take a different tact.

    The most comprehensive classification approach is the short paper by Don White [13, 15]. He defines six categories of classification and presents a table for each: A. Measurand, B. Technological aspects of sensors (performance specifications), C. Detection means used, D. sensor conversion phenomena, E. Sensor materials and F. Fields of application. A silicon pressure sensor, for example, would be described as having "mechanical" as its means of detection; the sensor conversion phenomena (piezoresistance) is "elastoelectric" and the key sensor material is an inorganic semiconductor. It would also be described by 12 different performance characteristics (sensitivity, range, speed, etc.) and be designed for use in at least one of 12 broad fields of application. White's table D of "sensor conversion phenomena" is very

similar to Middelhoek and Audet's classification by signal conversion [2,3]. (They arrived at this approach independently and nearly simultaneously; they don't cross-reference each other's work.)

Middelhoek and Audet's book [3] is restricted to silicon-based sensors but is very thorough. It summarizes three main approaches to the classification of sensors: by field of application (process control, biomedical, consumer, automotive, etc.), by measurand (temperature, pressure, optical image, presence, viscosity, etc.), and by measurement principal. His approach to the measurement principle is based on signal conversion from one of six domains (thermal, mechanical, radiant, chemical, magnetic or electric) into the electrical domain. His approach is very basic and has clearly been influential in subsequent publications. He also points out that sensors can be classified in terms of their purpose; they can be information oriented (e.g., a traffic situation), measurement oriented (e.g., distance) or physical quantity oriented (e.g., light intensity).

Some authors classify sensors as being *active* or *passive*, depending on whether they require an external energy source to power the sensor. Unfortunately, as both Middelhoek [3] and Carr [17] point out, different authors use these terms with exactly opposite meanings. Middelhoek [3] uses instead the terms *self-generating* and *modulating*, which are unambiguous, and similar to the terminology we introduce below (*energy conversion* and *energy modulation*). Fraden [26] distinguishes between *direct* and *complex* sensors. A direct sensor generates an electrical signal directly by passive (self-generating) or active (modulating) means. A complex sensor requires one or more energy transducers before a direct sensor can be used to generate an electrical signal. The present paper can be viewed as a further development of classification concepts described by Middelhoek and Fraden. It also classifies for the first time, a large number of sensor products according to sensing mechanism.

In addition to publications that give scientists' and engineers' classifications of sensors, it is relevant to look at sensor classification from the sensor business point-of-view. Examples are available on several web-sites [29-34]. Another example is the worldwide analysis and forecasts of sensor markets from Intechno Consulting [28]. This report is available to subscribers that have included sensor manufacturers, some technical specialists and probably few academics. In order to cover the field of sensors in an organized fashion, the report classifies sensors according to the purpose for which a sensor product was intended. "Purpose" generally produces two classifications – by measurand and by application. The market analysis is aimed at the sensor manufacturer, and therefore there is a third classification that is less explicit based on sensor technology (silicon, fiberoptic, etc.). It is interesting that many more kinds of sensors are included in the sensor marketplace than are included in literature reviews of sensor principles and applications. An exception is Fraden's "Handbook of Modern Sensors" [26], the most complete reference known to the present authors. On the other hand, for a popular measurand such as pressure the sensor literature has far more technical approaches than have appeared in the marketplace. (Note: A thorough review of the literature listing 110 books, 18 journals and 83 papers is given by Middelhoek [27].)

## 4. CLASSIFICATION BY SENSING MECHANISM

This paper proposes a new methodology for describing and classifying sensor mechanisms, namely in terms of ETPs. The intent is not to replace other sensor classification schemes, but to examine in more depth the basic mechanisms of sensing. Our classification scheme is based on the way information is gathered and the way it flows through the sequence of ETPs that comprise a basic sensor. We posit that ETPs can be classified in four distinct types as described in Box 1 below, at least for technology-based IT sensors. We test this hypothesis by applying it (in Table 3 below) to about 100 types of commercial sensors along with an estimate of the 2003 global market for each sensor type. We also apply it to information gathering devices and systems that are not usually labeled as "sensors" (see Table 4 below).

The purpose of classification is to identify common characteristics of basic sensors and sensing subsystems as information transducers, not to describe their mechanisms in detail. For detailed discussions of sensor mechanisms and technologies we refer the reader to the various comprehensive reviews of sensor mechanisms and applications [10-27]. Unfortunately, such reviews do not give the reader a sense of the whether a particular technical sensor approach is a well-developed product or is still at the development stage. This can be done by going to web sites maintained by the sensors industry [29-34] but these in turn do not indicate the relative importance of each sensor type in terms of market size. This information is only in a comprehensive market study such as the Intechno Report [28]. Based on information in that report, we give (in Table 3) the global market for each sensor category and identify the mechanism classification for the majority of sensors in that category.

The proposed classification scheme for transducer mechanisms is based on identifying the ETP type for each step of the information flow through a particular sensing subsystem. The four types of ETPs are:

1. Energy conversion (C),
2. Energy dispersion (D),
3. Energy modulation (M), and
4. Modulation of a physical property (P).

These are defined with brief examples in Box 1 and are discussed in detail in section 5 below. The classification process is based on the ETP sequence for a given sensor and is illustrated in boxes 2-4, with a wide list of examples in Table 3 below. The familiar p-n junction photodiode is a sensor that consists of a single ETP, so it is simply classified as Type C (energy conversion) or Type M, depending on the bias conditions. A lens is an example of a Type D sensor; it must be followed by a detector array in order to detect an optical image. An optical barcode scanner is an example of a Type M sensor that uses a laser probe that is modulated by reflection from the black and white stripes of the barcode. A thermistor is a very simple sensor but it requires two ETPs since two independent processes are involved. The first is the change in temperature induced by the outside world, changing the resistance (Type P), and the second happens only with the application of voltage (or current), which acts as a probe of the resistance (hence Type M). The two processes take place in the same device (the thermistor) so that the classification can therefore be described as Type PM.

Boxes 2, 3 and 4 illustrate this classification process for a pressure sensor, a video camera and a bicycle odometer. These very different sensing systems illustrate information flow in a sensor and the process of identifying the ETP types. The silicon pressure transducer (Box 2, Type C-PM) is familiar to all sensor technologists, but it may come as a surprise that energy conversion is involved in its sensing process. The video camera (Box 3, Type M-D-D-C) provides an example of Type D and M transducers. It also illustrates how some elements are part of the sensing subsystem but not part of a sensor product. The bicycle odometer (Box 4, Type C-C-C) provides an example of a high-performance low-cost instrument involving a series of 3 Type C transducers. The next four sections describe the four ETP types in more detail and with more examples, followed by a section that includes a table classifying about 100 commercially available sensors according to measurand and mechanism. It also provides a sense of their relative economic importance in the sensor marketplace.

**Box 1. Information transduction processes.** The flow of information through a sensing subsystem is a sequence of identifiable steps which we call elemental transduction processes (**ETP**s). Each ETP is characterized by certain features:

- Information is transformed from one type of signal to another.
- A characteristic relationship between energy and information flow.
- A transfer function relating output to input, response-time behavior, distortion, nonlinear behavior.
- Selectivity characteristics describing its ability to reject undesired inputs
- A characteristic noise behavior.

We posit that the following four types of ETPs form a complete set for describing information transduction mechanisms. The symbols are explained in Section 5.1.

**Type C** - **Energy or information conversion**. 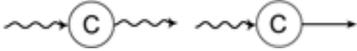 The incoming energy contains the desired signal and is converted to a different form of energy. **Strengths:** Lack of zero drift, wide dynamic range, simple calibration, excellent stability and resistance to environmental influences. **Weaknesses:** Sensitivity is limited by the measurand energy available. Many measurands have no available energy and cannot be sensed by this means. **Examples:** a photodiode converts optical energy to electrical current; a thermocouple converts a temperature difference (thermal energy) into electrical energy; a piezoelectric transducer converts mechanical energy to electrical and a clamped diaphragm converts pressure changes in a fluid to mechanical potential energy in a solid.

**Type D** – **Energy or information dispersion**. 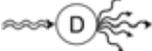 The incoming signal is separated (or dispersed) into two or more components. Implies a separable signal. **Strengths:** Same as Type C. **Weaknesses:** Must be followed by another stage of transduction to obtain an electrical output signal. **Examples:** Prisms, gratings (wavelength dispersion). Lenses (redistribute energy at an image plane). Optical filters (transmit narrow-band light and reflect other wavelengths). An (ideal) gas-permeable membrane passes a selected analyte gas and is impermeable to other sample gases. Gas and liquid chromatography columns, electrophoresis, mass spectrometry.

**Type M** – **Energy modulation**. 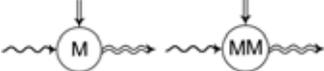 The incoming signal modulates a probe energy or signal. The probe (or carrier) source is part of the sensing subsystem and contains information about itself, but no information about the measurand. **Strengths:** Enables readout of stored information and other measurands that cannot be measured by Type C. Sensitivity is limited by the energy of the probe not the energy of the measurand (as in Types C and D). Type M transducers are especially useful in applications requiring high sensitivity where sophisticated signal processing can be used. Multiple probes may be used in the same device. **Weaknesses:** Requires an energy source. The probe energy may perturb the signal. **Examples of probes**: Optical, ultrasonic, RF, radar, magnetic and electric. **Measurands**: Distance, presence/absence, images (See Box 3), stored information.
Alternatively, energy is used to modulate the signal. **Modulators:** Choppers, injection valves.

**Type P** – **Property modulation**. 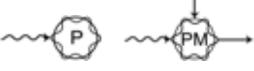 The input signal modulates the physical dimensions or material properties (or both) of the transducer material. **Strengths:** The only choice for many measurands such as static temperature and static pressure. **Weaknesses:** Type P transducers can have problems with baseline drift and stability. A Type M process must follow them to produce an output. The P and M processes may be in the same device as indicated by the second symbol. **Examples:** Thermistors, piezoresistors, magnetoresistors, color changes.

**Box 2.** A pressure transducer [45] illustrates several aspects of today's sensors. 1) It is a "smart sensor" that includes signal processing and signal processing features. 2) It is a silicon-based MEMS sensor, batch-fabricated by micromachining of a silicon wafer. 3) It illustrates the flow of information as it is transformed from fluid pressure (the measurand) to a precisely defined digital electrical output. The first information transducer (Jones' "primary sensing element" [12]) is a micromachined diaphragm that deforms in response to the difference in pressure difference, $\Delta P = P_2 - P_1$, on the two sides of the diaphragm, C. The diaphragm is Type C because the external pressure does work to deflect the diaphragm (converts gas pressure to mechanical potential energy). It is a highly selective mechanism. Only the *change in pressure* does mechanical work, changing the stored potential energy and the strain pattern in the stretched diaphragm. The outer region of the diaphragm has built in piezoresistors whose resistance depends on strain (**Type P**). The next step of information transduction is the result of the voltage applied across the Wheatstone bridge, a modulation (**Type M**) process. The Wheatstone bridge is a single device that combines the two processes, symbolized by the merged symbol, PM. The information flow is: $\Delta P \Rightarrow$ Strain $\Rightarrow$ Resistance change $\Rightarrow$ Voltage $\Rightarrow$ Signal Conditioning (SC). A similar mechanism applies to the temperature (T) sensor used for temperature compensation. The "smart" electronics include stored data from the calibration process, compensation for zero offset, temperature and pressure nonlinearity and outputs in different formats for either analog or digital controllers and displays.

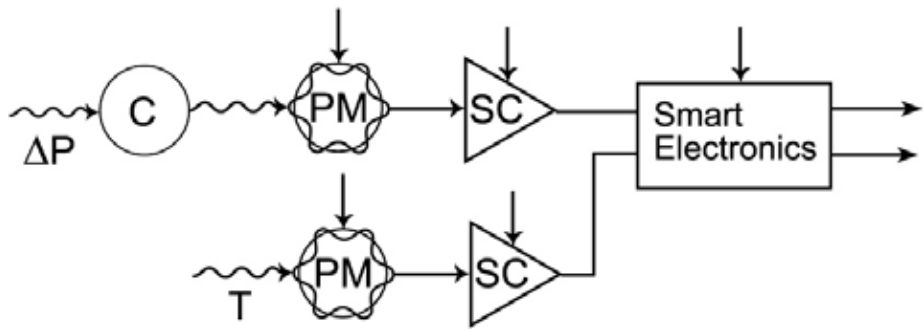

**Sensor classification: C-PM** for the pressure sensor, **PM** for the temperature sensor.

**Box 3.  Color video camera.**

The familiar color video camera is a high performance technology-based sensing system.  It provides high quality data at a high data rate with a high degree of selectivity – it is not sensitive to environmental variables such as temperature and humidity (within reason).  It also has a very wide dynamic range – able to operate in candlelight or in bright sunlight.  We commonly think of the imaging chip as the sensor because it provides the electrical signal output.  However, the sensing subsystem includes much more.  The first transduction of information occurs at the surface of the object where light is reflected, a **Type M** transduction process.  The energy source can be the sun, which provides uniform illumination coming from a certain direction, with a characteristic intensity and spectral distribution.  It carries information about itself (direction, intensity, spectrum) but not about the measurand (the object).  The illumination source and the object (scene) are essential to the sensing mechanism but certainly not part of a sensor product.  The input signal, S, is the information stored in the scene and the outgoing signal is the spatially modulated light, $S_1$.

The second ETP is the lens $D_1$ that focuses the signal $S_1$ to each element of the pixel array and performs the critical function of selective omission of information: it defines the field of view and the resolution (spatial frequency) of the image. Color filters $D_2$ separate the signals into three components.  Each photodetector (**Type C**) in the CCD array converts the optical flux to an electron current, which charges a built-in capacitor.  The charge stored in each capacitor is then read out sequentially to generate the video signal.  This readout, also a modulation process, occurs after the conversion to an electrical signal and thus is not part of the basic sensing mechanism.  If the video camera is part of a security system it has a motor-driven panning mechanism that selects which parts of the scene are viewed at which time.  Panning, auto focussing and iris control are auxiliary functions important to system operation but they are not in the direct information flow path.

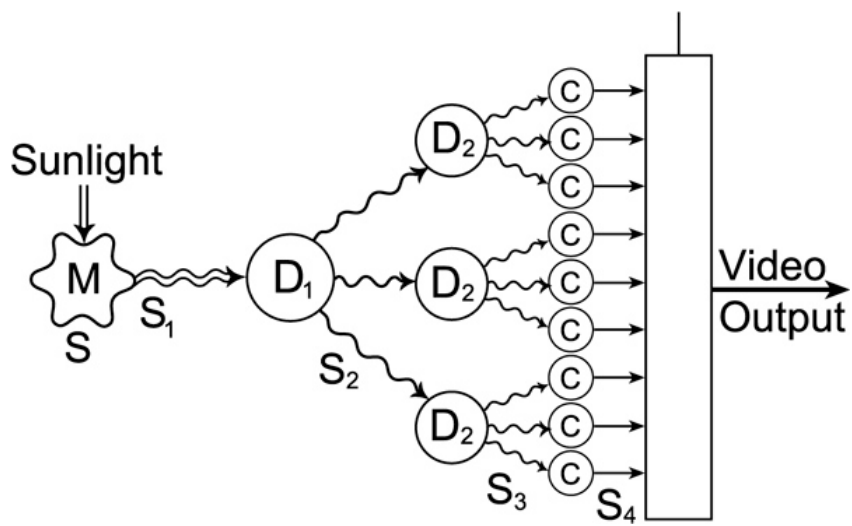

**Sensing process classification: M-D-D-C**
Sensor product consists of: C only (assuming a packaged CCD array; in a CMOS array M would replace C)

**Box 4. Bicycle odometer.**

A bicycle odometer/speedometer is an example of a sensing system that illustrates how high quality sensor information can be provided at a low cost, by converting the signal to digital form as early in the chain as possible. According to our marketplace definition of sensor it is not a sensor, because it does not have an electrical output. The marketed product consists of the magnet to be attached to the wheel, the pick-off coil to be mounted on the frame and the electronic package that includes the display. The bicycle wheel is not part of the sensor but is certainly in the path of information flow and thus is part of the sensing subsystem. The energy for the sensing process is provided by the bicyclist (who is also not part of the sensor).

The measurand signal of interest, S, is linear displacement. The front wheel, $C_1$, converts this signal to $S_1$, rotational displacement. Linear kinetic energy is converted to rotational kinetic energy. The magnet mounted on the wheel $C_2$ converts the rotational motion to a time varying magnetic field $S_2$ and a pick-up coil on the front fork $C_3$ converts the magnetic pulse to a voltage pulse $S_3$ based on Faraday's law of induction. The three transducers are all **Type C** transducers, converting some of their incoming energy to a different form of energy. The voltage pulse is the trigger for the digital counter that counts the number of revolutions the wheel makes and provides a distance measurement. Its accuracy is determined by the accuracy of the wheel diameter measurement, which is entered in as a calibration parameter. The odometer also uses a quartz clock for velocity calculation.

The clock consists of a piezoelectric quartz resonator with a very high mechanical Q. The piezoelectric resonator is a narrow band filter (**Type D**) that is driven electrically (**Type M**) and combined with an electronic amplifier (A) in a closed loop circuit form an oscillator. (See also Section 5.2.) The microprocessor calculates distance, speed and average speed in the desired units.

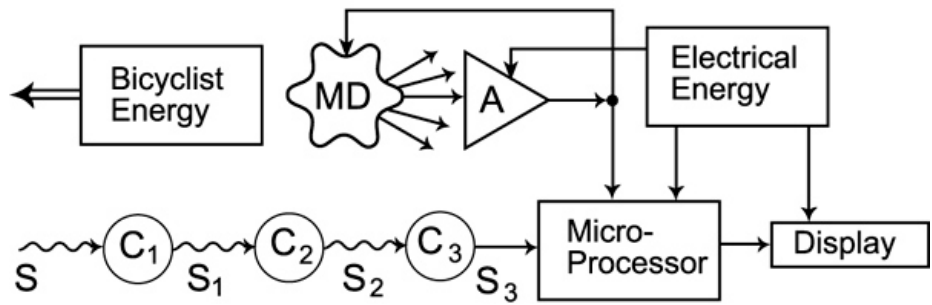

**Sensing subsystem classifications: C-C-C** for the distance sensor, **MD** for the clock. Product consists of kit containing magnet, pickup coil and electronics package.

## 5. FOUR TRANSDUCTION MECHANISMS: C, D, M and P

This section describes in more detail the four types of elemental transduction mechanisms introduced in Box 1. The new classification system consists of identifying the sequence of ETP types (C, D, M and P) that describe the information flow from the external world to an electronic signal. Although strictly speaking the four types refer to processes, it is convenient to use the same terminology for the transducers (devices, structures) that implement those processes.

### 5.1 Transducer symbols

As illustrated in Boxes 1 through 4, the understanding of sensor mechanisms is aided by the use of symbols to represent the four ETPs. Information flow is from left to right. Wavy lines represent non-electrical signals and straight horizontal lines represent electrical signals. Vertical lines indicate auxiliary energy inputs that do not contain a signal. Circles represent transducers with a letter (or letters) denoting its mechanism (ETP type). A wavy circle represents information related to a property or structure variation, such as information stored in an object that can be read by an optical probe. A circle with a superimposed wave represents the result of a Type P process – a time-variation. A Type M must always follow a Type P process because property changes by themselves do not provide output signals. This suggests the merged PM symbol. The piezoresistive bridge and thermistor are examples (Box 2). There is only one device (the bridge) but there are two transduction processes.

The idea that more than one process can be taking place in a single device is hardly new. An example is the local oscillator and mixer that occur in the first stage of a superheterodyne receiver. The two functions are combined (or merged) in a single transistor. Another example is the high-Q piezoelectric resonator in a quartz clock (Box 4) represented by merged MD symbol. The resonator is driven by an ac probe signal (Type M) provided by a broadband amplifier but it responds selectively (Type D) to a very narrow band centered on its resonant frequency. The information about the resonant frequency is contained within the structure and is represented by the wavy circle.

We have illustrated ways in which the number of transduction processes exceeds the number of transducer elements. One way in which this can happen is with the use of multiple probes of the same property change. For example the SQUID mechanism that involves an rf probe and a dc probe [35]. The atomic clock [36] and sophisticated instruments such as the ones listed in Table 4 below also make use of multiple probes.

### 5.2 Energy conversion (Type C)

Energy conversion transducers convert the energy of the incoming measurand to a different form of energy. The most-cited examples are those that convert energy to electrical energy. Piezoelectric transducers, for example, convert an acoustic or vibration signal to an electrical output. These transducers also can function as output devices because they obey reciprocity; mechanical energy can be converted to electrical energy or vice versa. Many other Type C transducers do not obey reciprocity: photodiodes, thermocouples, RF antennas and magnetic pickups that use a changing magnetic field to generate a voltage by making use of Faraday's law of induction. Antennas for converting electromagnetic waves into electronic signals in radio, TV and GPS systems are not considered sensors, but are Type C input transducers, according to the above definition. Electrochemical sensors such as the oxygen sensor in our automobiles convert chemical energy (actually the chemical potential difference due to a difference in oxygen partial pressure) into an electrical signal. A magnet on a rotating shaft generates a varying magnetic field in an ignition-timing sensor. The bimetal strip in a thermostat is an example of a Type C transducer that does not (directly) have an electrical output; it converts thermal energy to mechanical energy.

Many sensors make use of Type C transducers that have a non-electrical output signal; they must be followed with a transducer with an electrical output (another C or an M). Examples include capacitive sensors, spring proof-mass combinations (in accelerometers), load cells for force and weight measurement, diaphragms in pressure transducers and in microphones. Whereas Type C transducers can measure *dynamic* pressure (as in sound wave), *static* pressure cannot be measured by a Type C mechanism alone since no energy is available for conversion. A piezoelectric transducer is a type C mechanism well suited for vibration measurement but even in the charge read-out mode it cannot measure static pressure because there is always finite charge leakage. However, a Type C transducer can cause a material property to change as illustrated in Box 2. While some electrochemical sensors generate an electrical output directly, others such as the blood glucose sensor convert one chemical to another and thus make use of the Type C mechanism.

Silicon photodiodes are perhaps unique in that they can be operated as either Type C or Type M devices. If their output is fed into a low input impedance signal processor (such as a current-to-voltage converter) they are Type C transducers; if they are reverse-biased with a high impedance load resistor they are Type M. In either case the transducer *is* the sensor; only one stage of transduction is involved. They make use of the internal photoelectric effect, the well-known energy conversion mechanism used by solar cells. In well-designed silicon photodiodes each photon of visible light produces an electron-hole pair so that light intensity can be very efficiently (>99%) converted to electron current. Photovoltaic transducers illustrate the strengths of Type C processes mentioned in Box 1:

1. There is no zero correction or zero drift. Zero incident photon flux in means zero output current, by conservation of energy. As an energy conversion device it responds only to optical energy in its specified wavelength region. Any zero offset is due to the electronic circuit at the transducer output.
2. Calibration is very simple, because the main variable in the process is the area of the device, which is determined by precision lithography of the type used in integrated circuits. The energy conversion mechanism does not depend on temperature, thus calibration does not need to include temperature effects. There is essentially no interference from magnetic fields, vibration or other unwanted effects.
3. The dynamic range is very large. The photon-to-electron conversion efficiency can be maintained over a very wide range of intensity levels – up to 8 orders of magnitude. In fact, specially designed silicon photodiodes provide a secondary standard for the measurement of light intensity.

In addition, silicon photodiodes can take advantage of silicon technology. They can be fabricated in arrays and read out sequentially using electronic circuits fabricated in the same silicon chip. Such CCD or CMOS arrays are the basis of today's video cameras and digital cameras as well as astronomical telescopes.

### 5.3 Energy dispersion (Type D)

Type D transducers disperse and/or select the incoming signal, separating it into multiple channels without energy conversion. This transduction process only applies to signals that have multiple components. Four different kinds of dispersion have been identified: wavelength, spatial, time and chemical selectivity. The name "dispersion" comes from examples of optical transducers: wavelength dispersion in prisms and gratings. Such Type D transducers are at the heart of spectroscopic analysis. A lens is a Type D transducer when it disperses incident light to the pixel elements of a detector array, forming an image at the detection plane. Spectral imaging or color video requires two successive Type D processes or transducers (See Box 3): the objective lens and a band pass filter system for separating white light into the desired three colors.

Chemical sensors and biosensors depend heavily on Type D transducers, because they have the challenge of detecting particular analytes in a mixture of other chemical species. Sensors for a particular gas or particular ion in solution make use of gas permeable membranes and ion-selective electrodes, respectively. These transducers separate on the basis of chemical selectivity, ideally selecting one analyte species and rejecting all others.

Instruments for analytical chemistry use sophisticated Type D separations. Examples include: the gas chromatography, which separates by selective adsorption/desorption on the basis of boiling point; electrophoresis, which separates on the basis of ion mobility in a liquid or gel; the ion mobility spectrometer, which operates on a gas mixture and the mass spectrometer which separates on the basis of mass-to-charge ratio. These analytical instruments are based on time-of-flight separation; an injected pulse of sample is dispersed into separate pulses, each corresponding to an individual analyte. The pulsed injection mechanism generates the pulse modulated output signal, a Type M process as discussed in the next section. MEMS technology is currently being developed to realize these powerful sensing mechanisms as compact unattended sensors and in control systems [37].

### 5.4 Energy modulation (Type M)

Modulators are devices such as shutters and chopper wheels that modulate a signal such as an optical beam. A shutter in a thermal infrared imager presents the same temperature to each pixel in the detector array, providing a reference for comparison with the temperature in the scene. Injection valves perform a modulation function when they inject a chemical sample into a reference gas stream in a gas chromatograph as mentioned above. The purpose of modulation is to separate the signal from a background and thus is a powerful signal-processing tool.

The classic example is the chopper wheel in thermal imaging infrared detectors. This is a segmented wheel that mechanically chops the incoming light so that the detector alternatively sees the scene and the chopper. The chopped signal is an ac signal with the temperature of the chopper blade as reference. The ac signal is detected by phase-sensitive electronics, avoiding the problem of zero drift in a PM type of infrared detector (such as photoconductive detectors).

A similar type of chopping enables chemical analytical instruments that use time dispersion to separate analyte species. These involve Type M-D transducers as mentioned above because the pulse injection modulator (Type M) must precede the Type D element in each case. In a gas chromatograph, for example, the modulator is the valve that injects a pulse of multi-analyte sample into the carrier gas that then flows through the separation column to subsequent transducers that produce the electrical output. The electrical output is a pulse-modulated signal with each pulse corresponding to a single analyte. The time of a pulse measured from the time of the injection pulse is called the separation time and is the identifying characteristic of each analyte.

But modulator devices are not the only way in which a modulation process can be used for sensing. The signal of interest can be used to modulate an energy probe, thereby converting the signal to modulated energy. The modulated output is a measure of the measurand. The energy source may be clearly external; it may or may not be considered part of the sensing subsystem. As we shall see in Table 3 below such Type M processes are the most prevalent sensing mechanisms found in commercial sensors. A simple example is an automatic door opener based on the interruption of an optical beam. A powerful (but less obvious) example is the transduction process associated with optical imaging illustrated in Box 3. Imaging requires an illumination source, such as the sun in the case of an outdoor scene. The reflection of light from the object is a modulation process in which the (uniform) illumination is transformed into light carrying information about the object. The illumination source contains information about itself (spectral content, direction) that contributes to the reflected information, but does not have measurand information.

Laser scanners (bar code readers) at the checkout counters of grocery stores, in copying machine and fax machines and optical read heads in CD and DVD players illustrate the possibilities of Type M transducers based on optoelectronic technology. The light from the light source is coded by wavelength and intensity modulation so that the sensor is not influenced by stray light, which would otherwise flood the photodetector. Other examples include fire detectors based on light scattering, optical encoders in speedometers and tachometers, optical distance sensors, turbidity sensors and surface roughness sensors. Some of these use LED light sources and others use p-n junction lasers, but they all use the Type M mechanism.

Some scanning systems use the motion of the object rather than a scanning probe. This is particularly true for stored magnetic data: credit cards and room keys that have to be "swiped," magnetic tapes, disks and hard drives, for example. The movement of the object provides the energy that produces the modulated output. Therefore, the mechanism is consistent with the definition of Type M. In the case of stored magnetic data the changing magnetic field produced by the motion provides the energy for readout by a pickup coil, an energy conversion (Type C) transducer. Magnetoresistive read heads also require AC magnetic fields even though they use a magnetoresistive (Type P) mechanism, because they are limited by 1/f noise at low frequencies. In an optical encoder where the light source is part of the sensor head, the motion of the black/white stripes on the reflecting object produces the light modulation.

Sound, ultrasound and microwaves are other forms of energy used in Type M transducers. The sonar systems used by bats and dolphins are nature's examples of such sensors that illustrate the principle. The modulation can take the form of time-of-flight measurement to determine distance or it can use Doppler effect (the frequency shift compared to the source frequency) to measure velocity. These illustrate the selectivity and sensitivity achievable with Type M sensors because of the possibility of control over the energy source.

Sophisticated imaging systems enabled by today's technology also use type M transducers. X-rays, electron beams, positrons, and other particles can be used as probes to form images of surfaces and identify the atomic nature of a surface. The object being examined modifies the beam and provides a unique signature to a detector. Modern medicine depends heavily on this technology and the images are an important part of the data on each patient. Material science has also come to depend on these types of instruments. The electron microscope forms images of surfaces at magnifications that are not possible with optical microscopes. Auger microscope augments the electron microscope's capability with the ability to identify the chemical composition of the surface with micron-scale resolution. Atomic force microscopes and scanning tunneling microscopes go to the atomic scale, with the ability to image individual atoms on a surface. All of these instruments depend heavily on information processing, but the information is gathered by a sensing mechanism that clearly falls into the Type M classification.

It is interesting that certain measurands and other input parameters require transduction by a Type M process. Distance, spatial dimensions and surface structure are examples. All transducers for imaging, whether scanned or imaged with a lens and all transducers for reading stored data are also Type M. These measurands do not have any energy associated with them that could provide input to a Type C or Type D transducer nor to produce changes in the material properties for Type P transduction. An auxiliary energy source of some sort is necessary.

As previously noted, photodiodes can be operated as either Type C or Type M mode. Many photodetectors use current-to-voltage converter circuits that keep the voltage across the diode as near to zero as possible to insure the advantages of the Type C operation over a wide dynamic range. With reverse bias they become Type M devices with current proportional to light intensity and supplied by the voltage source, not the photons themselves. In silicon the offset (or dark) current is nonzero but very small, and the dynamic range is quite limited, but the reverse bias mode provides signal amplification that can be cost-effective in many applications. It is quite similar to amperometric operation of an electrochemical cell. In both cases, the device is operated in a plateau region where the current is almost independent of voltage, and the current carriers (electrons or holes in the photodiode and analyte ions in the electrochemical cell) are collected with high efficiency. Photoconductors have no such plateaus and operate by modulating the resistivity of the material, making them Type P devices.

### 5.5 Modulation of material properties (Type P)

Type P transducers are well suited for many measurands and are required for some that cannot be measured by the other three types. For example, static temperature must be measured by a Type P mechanism, while thermocouples and pyroelectric detectors can measure *temperature differences* and *temperature changes*, respectively by a Type C mechanism. A property change does not produce an output signal so that generally a Type M transducer must follow a Type P transducer. A very simple example of Type P temperature sensor is a thermistor or a platinum resistance thermometer. The resistance is measured by applying a voltage and measuring a current – a Type M mechanism.

Resistance change is frequently the property measured in Type P transducers: piezoresistors for strain, chemistors, humidity sensors, magnetoresistors in various flavors (AMR, GMR, CMR and EMR) [23, 38] and the Figaro gas sensor [39] for combustible gases such as methane. These are discussed in Section 17 below. Other material properties that can be modulated by a measurand are channel conductance, membrane permeability, membrane potential, and enzyme activity.

The fact that resistance change is frequently used as a sensing mechanism illustrates one of the problems with Type P transducers. Electrical resistivity can be intrinsic to the material, extrinsic (dependent on impurities in the material) or can be due to surface properties (surface conductivity). It can therefore be influenced by many variables and thus is subject to interference from signals other than the desired measurand. It other words, Type P transducers tend to suffer from poor selectivity and baseline stability, the bane of sensor developers. This weakness can be ameliorated by the use of a signal modulator as discussed above.

An unexpected feature of this classification methodology is that "capacitive sensors" can have different classifications. If the capacitance change is due to a dimensional change (the plate spacing in an air-filled capacitor, for example) the transducer is Type C, since the measurand energy must do mechanical work (on a spring, for example) to change the dimensions. No modulation of material properties occurs, although material properties (Young's modulus) are involved. On the other hand, if the change is due to a change in dielectric susceptibility (due to a liquid filling the capacitor, for example) the transducer is Type P. Both cases require a Type M transducer such as an impedance bridge to measure the change in capacitance. The Type C capacitive transducers are more selective than Type P, because they can be designed so that the plate spacing changes only in response to the work done by outside pressure. An example of the potential probes with Type P is the aging of a capacitive humistor based on water molecules diffusing into the dielectric between the plates. Its impedance will decrease over time as a result of ionic contamination from the environment.

The separation of dimensional changes and shape changes (Type C) from material properties changes (Type P) is also appropriate from a manufacturability viewpoint. The control of dimensions in the manufacturing process is much different than the control of material properties. For example, when silicon is used as a mechanical material such as in a cantilever spring in a MEMS accelerometer its deflection in response to acceleration is a function purely of its dimensions, which can be controlled by precision micromachining (photolithography and etching). Of course, it also depends on Young's modulus, but this is a bulk property of the single crystal material that is insensitive to doping and is highly reproducible. Sensor material properties, such as the resistance of the semiconductor materials or metal alloys used for sensors (piezoresistors, thermistors, etc.) may vary widely and depend critically on the fabrication process.

## 6. SENSING MECHANISMS ASSOCIATED WITH COMMERCIAL SENSORS (TABLE 3)

A central theme of this paper is that the four ETP types described above form a complete set, i.e., they identify the basic mechanisms of all IT input transducers. Proof of this hypothesis would require examination of all known sensing mechanisms on a case-by-case basis, which would be an overwhelming task. Instead, consideration has been limited to the wide variety of sensors and information transducers used in today's IT systems: those products that are sold as sensors on the open market. The intent is to list them in terms of their intended purpose as specified by sensor manufacturers. The market categories chosen are from the 1998 Intechno sensor market study [28], with minor changes. They are listed in Table 2, which serves as a table of contents for Table 3. Data input devices and instruments that are not considered parts of the sensor market are considered in the next section and in Table 4.

Usually purpose translates to measurand. The nine categories in Table 3 correspond to measurands and are in rough agreement with the classifications in other references listed above. In each category binary sensors (those designed to have a binary output) are grouped together as the first subcategory. Each entry in Table 3 then describes a sensor technology based on a specific mechanism (or a closely related mechanism) and is characterized by a certain market size. The 2003 global market for each line item was estimated from the 1998 market data. A complete analysis of sensor markets by country, by technology and by industry with projections to 2008 is given in the full report [28]. An advantage of a market-based classification is that it enables an assessment of the relative economic importance of different sensor technologies. This classification approach is not without difficulties, however. Technologies for the same measurand but used for different purposes are listed under different categories. Thus magnetoresistance sensors are basically magnetic field sensors, but could be used for proximity sensing, rpm, engine timing, compass direction, reading stored data as well as magnetic field strength, which are all quite different purposes and markets with different competitors in each market. Looking only at the markets under category 7 - "Electrical and Magnetic Quantities" would significantly underestimate the market. Other examples are discussed below as they arise in each category.

The ETP mechanisms listed in each line of Table 3 refer to the sensing subsystem associated with the sensor product. As we have mentioned before, a sensing mechanism can correspond to a range of sensor products, and all are included in the same line of Table 3 if they make use of the same sensor technology. Each identifiable sensor technology (materials, method of manufacturing, etc.) enables certain ETP combinations and measurands and defines their performance limitations. One manufacturer may be strong in devices based on certain types of structures at low production volumes, while another may be strong in a wider variety of products but require large production volumes to be profitable. Pressure sensors provide an example. Although they use the same mechanism, strain gage technology is quite different from MEMS-based technology. Large, low cost diaphragms can easily be stamped out of metal, but the attachment of foil strain gages and clamping the edges become the challenge. MEMS micromachining is a versatile batch manufacturing processes in which each diaphragm is integrated with the piezoresistors and perhaps even a vacuum reference chamber. Similar MEMS processes can also be used for flexures and proof masses for capacitive accelerometers.

### 6.1 Binary sensors (all categories)

Each of the nine measurand categories in Table 3 starts with a subcategory called binary sensors. Binary sensors provide only one bit of information: the state to be detected has been achieved or it has not yet been achieved. We list them separately because sensor manufacturers distinguish the market for binary sensors from those intended for measurement, even if they use the same technology as sensors intended for measurement. They are not mentioned as a category in other sensor reviews, but are discussed here in some detail because they illustrate recent trends in sensor technology, as well as some of the difficulties in classifying sensors.

Binary sensors are important because in many applications the desired outcome of the sensing process is a binary output. For example, in many (perhaps most) control systems binary control is the most cost effective because adequate control and maximum efficiency can be obtained by time modulation of the output through closed-loop operation. Residential controllers such as furnaces, air conditioners or ovens are either on or off. The high efficiency of hybrid automobiles is another example. Binary outputs are also desired for alarms or status indicators that do not have a control function.

The simplest and oldest binary sensors, electromechanical limit switches, are the first entry in Table 3, even though they are not considered part of the sensor market. The market number in Table 3 is approximate. "Touch-based" or tactile switches for computer input are included as category 11 in Table 4. Even though electromechanical switches are usually not classified as sensors, Fraden [27] recognizes their significance when he describes the reed switch as a binary proximity sensor. The reed switch consists of a pair of hermetically sealed spring contacts made of a magnetic material. The contacts are pulled together in the presence of a magnetic field. The mechanism is Type C since external work is done to pull the contact together. Reed relay technology (a reed switch with a solenoid wrapped around it) has been very important in the past because reed switches represented a cost-effective way of implementing telephone switchboards before the invention of the transistor. Sensors that detect presence or absence of objects are also considered binary position detectors. It is a sign of the times that the technology is evolving away from simple electromechanical switches toward devices that are "touchless," i.e., proximity sensors. Today most binary position sensors are touchless with manufacturers of analog position sensors selling binary proximity sensors based on the same technology. The Hall switch is a solid state device that gives a binary output in response to the proximity of a magnet [3,18,40]. Induction-based proximity switches are actuated by the proximity of a metal conductor. The hysteresis inherent in mechanical switches is a desirable feature that reduces noise and is implemented electronically using a Schmitt trigger circuit [27]. Such solid state sensors are replacing mechanical limit switches because they have the advantages of:
- Avoiding electrical contact problems due humidity or corrosive environments,
- No mechanical actuation force is needed,
- High frequency operation,
- Long lifetime – no mechanical wear, and
- Immunity to shock and vibration during operation.
- No noise due to "contact bounce."

The higher reliability justifies the higher prices for the newer technology.

Some binary sensors are called detectors rather than sensors, e.g., metal detectors (Category 1.0), smoke detectors (Category 7.0) and glass breakage detectors (Category 5.0).

Their purpose is to detect whether or not an event occurred rather than to make a measurement. Binary position sensors and motion sensors (both Category 1.0) have the largest markets and include optical, infrared and ultraviolet light barriers, reflector-type photosensors and light curtains. Also in this category are magnetic-based, capacitive and ultrasonic proximity switches, such as proximity switches based on the Hall effect, magnetic induction loops for traffic infrastructure and radar sensors that recognize specific targets. The two basic mechanisms for smoke and fire detectors (Category 7.0) are optoelectronic (photoelectric) and ionization chamber (See Allocca [11], chap 20 and Fraden [27], chap. 14), both of which use Type M mechanisms. Rain sensors for automatic control of windshield wipers are also in Category 7.0 and are mainly based on optoelectronic (Type M) probes.

Airbag sensors (Category 1.0) are binary sensors that use the same basic mechanisms as acceleration/vibration sensors (Category 1.4). Airbag sensors generally are Type C sensors since the rapid deceleration causes an effective force that does work to displace a proof mass. The displacement can be sensed in a number of ways (Nwagbozo [18], chapter 8): electromechanically, piezoresistively and piezoelectrically. The piezoelectric accelerometer consists simply of a piezoelectric ceramic bonded to a proof-mass and is Type C-C. Recently capacitance-based MEMS technology has gained market dominance. Their advantage is that the capacitor and proof mass can be co-fabricated with a number of "smart" electronic features in the same CMOS chip [21,41]. They thus can be fabricated and packaged using the well-established CMOS integrated circuit technology base. With refinements of the technology they can be used for higher value applications such as navigation.

Binary sensors follow the general sensor trend toward miniaturization; more and more sophisticated electronic functions (Figure 1) are integrated into the smallest possible space. Surface mount technology, thick film hybrid circuitry, ASICs and microcontrollers are increasingly being used. For reflector-type photosensors and light barrier proximity sensor the result is the reduction of control system errors caused by network failure, optical faults and the effect of dirt [28].

The classic example of binary temperature sensors (Category 3.0) is the household wall thermostat. They are based on the bimetal strip, a Type C mechanism for converting thermal energy into mechanical work that opens or closes an electrical switch (Type M). Thermostats are sold as control elements, and are not considered part of the sensors market; thus the market number in Table 3 is approximate. Even though integrated circuit technology has advanced with dramatic reductions in cost, it has not been easy to replace the many subtle advantages engineered into the electromechanical devices. However, the many advantages of the digital technology and programmability are changing the marketplace so that thermistors, RTDs and diodes (Type PM temperature sensors) are rapidly gaining market share over the classic C-M mechanism.

We note that many binary sensors are based on Type M mechanisms. This makes sense in terms of the attributes (Box 1) of this ETP type. Only one bit of output is needed but the information must be robust. The use of an energy probe enables a large signal-to-noise ratio with high selectivity against undesired interferents. The airbag switch, for example, depends on a large change in capacitance that is not upset by normal deceleration inputs. The threshold can be set when the ignition is turned on and acceleration is zero; the precise value of capacitance is not important, in contrast to (analog) capacitance acceleration sensors (Category 1.4).

### 6.2 Sensors for mechanical quantities not related to fluids (Category 1)

Sensors for mechanical quantities fall into two categories, depending on whether they relate to solid bodies (Category 1) or fluids (Category 2). This is clear by from the subcategories in each category and illustrates the difficulty (or arbitrariness) of classifying sensors by measurand. The terms "position", "displacement", "motion", "proximity", "length", "distance" and "level" are closely related (Norton [14], chap. 5), but in the sensor marketplace can mean quite different devices. The terminology depends on the application and different applications sometimes make use of the same sensing mechanism. *Position* (Category 1.1) generally means the measurement of an object's coordinates (linear or angular) with respect to a specified reference. *Displacement* or *motion* refers to movement of an object from one position to another, and *Proximity* sensors generally refer to threshold position sensors; both are usually binary sensors (Category 1.0). *Level* refers to the position of a liquid surface with respect to a reference level and is listed under Category 2.3. Whatever name is used, such sensors are based on Type M mechanisms. *Tilt* or *inclination* sensors (Category 1.8) sense angular position with respect to the direction of gravity. Although they are indeed position sensors they operate on an acceleration principle as discussed below, and thus can be based on a Type C mechanism – the tilting process does work on the system.

The largest market for position sensors are based on optical encoders followed by magnetically-based resolvers or synchros. Their operation and application is described in Allocca [11], chapter 21 and also in Norton [14], chapter 5, who prefers the term "displacement transducers". The LVDT (linear variable differential transformer) principle uses ac excitation to establish a magnetic flux that is coupled through a symmetrically-wound secondary coil on either side of the primary (Allocca, chapter 8). Potentiometer-based sensors are used as well, mostly in the automotive and hydraulic business sector. There are two types of position sensors: incremental (or motion-based) and absolute. Absolute sensors have the advantage that the measured value is preserved when the power is interrupted due to network failure, system failure or pick-up interference. The development of absolute encoders shows the following trends [28]:

- Multiple efforts to reduce the number of mechanical elements (part count),
- Higher bit rates for robotic technology (18-bit encoders and higher) as well as other machinery (10-16 bit encoders),
- Lower priced encoders, driven by sewing machine and textile manufacturing applications,
- Serial interface to reduce cable complexity and cost,
- Encoders that are easy to sterilize for the food industry,
- Explosion-proof encoders using optical fibers for the chemical and petrochemical industries.

High-end encoders include highly integrated electronics with ASICs that provide programmability and PC interfaces. Shaft encoders flanged directly on a motor shaft can be configured to provide data on rpm as well as position so that a single encoder replaces the position controller, the tachogenerator and the Hall sensors used for brushless dc motors.

### 6.2.1 Navigation (position and direction) sensors (Category 1.2)

Navigation sensors perform distance and direction measurements on a much larger scale than position sensors. The field has become dominated by GPS systems, which are generally viewed as sensors in the nautical and aerospace industry and are therefore included along with other navigation sensors in the sensor marketplace. They are based on sophisticated signal processing of data from GPS satellites to infer latitude and longitude position. The position transduction mechanism is based on the timing of the satellite pulses, so the overall mechanism is Type M with the satellite electromagnetic signals providing the probe energy. This is consistent with the other position sensors noted above, which are all Type M. Interestingly, the primary sensing element of the GPS receiver is the antenna, which really is a transducer of EM energy to electronic signal and is therefore Type C. The classic needle compass is a Type C navigation direction transducer, but modern magnetoresistive compasses (Type PM) used in automobiles achieve the same result. Gyros of various types are all Type M with mechanical or optical probes. Ring laser gyros and fiber-optic gyros [42,43] are manufactured only as highly integrated IRS (Inertial Reference Systems) that include three gyros for the three axes of rotation as well as three accelerometers whose output is integrated to give position as well as direction. Their transduction mechanism depends on the principle of general relativity that enables them to detect rotation with respect to an absolute reference frame. Thus they can use the earth's rotation for calibration; their sensitivity can be less than 1/10,000 of the earth's rotation. Manufacturers and users generally refer to IRS systems as instruments rather than sensors.

### 6.2.2 Speed and RPM sensors (Category 1.3).

Rotation speed can be measured by a Type C mechanism (Box 4 above). DC-tachometers generate a DC voltage proportional to rotational speed; the polarity indicates the direction of rotation. AC tachometers that generate polyphase sinusoidal voltages are less expensive, but information about the direction of rotation is lost, and the linearity of a DC output is lost because of the threshold voltage of the rectification diodes. Optical digital tachometers or incremental encoders use a Type M mechanism to convert a rotation (rpm, angular position) optoelectronically into a sequence of low/high pulses. The number of pulses per turn determines the resolution.

An important application of rpm sensors is for control of servomotors. In the past AC and DC tachometers were perfectly suited for analog control systems. Digital control brings new requirements such as 10,000 measurements per second in digital format. This has given rise to a sophisticated generation of digital tachos (incremental encoders) and so-called sinusoidal digital tachos. In motor vehicles much lower priced rpm sensors are needed for ABS (anti-lock braking systems) and TCS (Traction Control Systems) systems as well as for gear control and speed indication. Various magnetic-based technologies (inductive pick-up coils, magnetoresistive heads and Hall effect switches) are used here (See Nwagboso [18], chapter 4). These are really "gear-tooth" proximity sensors that make use of a permanent magnet and the fact that automotive gears are made of ferromagnetic materials. Passive inductive sensors are the simplest but have trouble providing enough resolution at low speeds. Active Hall and magnetoresistive chip-based sensors with integrated signal conditioning sensors provide an output independent of speed. The trend is toward GaAs Hall-ICs that work at temperatures up to 150C. Such sensors are being integrated into ball bearing housings and provide output at near-zero speed. Their basic transduction mechanism is Type M-PM, with the magnetic material (ball bearing or gear tooth) modulating the magnetic field, that is then detected by a Hall (Type PM) sensor.

### 6.2.3 Acceleration/vibration sensors (Category 1.4)

Acceleration and vibration are sensed by similar mechanisms since vibratory motion must involve acceleration. Whether sensed by a piezoelectric, capacitive or piezoresistive mechanism, the primary transducer element is Type C - the acceleration of a proof-mass produces a force that does mechanical work on the sensor element. Except for piezoelectric sensors, a Type M mechanism is used to convert the resulting deformation to an electrical signal. Piezoelectric ceramics are traditionally the lowest cost and are the technology of choice, in industrial applications like paper mills, for vibration monitoring to indicate when maintenance is required in pumps, compressors and ball bearings.

The automotive industry requires acceleration sensors for airbags, automatic seat belts and electronic shock absorbers. These are based on piezoelectric, piezoresistive as well as capacitive mechanisms. Crash detection sensors have been included above as binary sensors (Category 1.0), but are generally based on accelerometer technology. MEMS-based capacitive sensors are increasingly displacing the other types because the electronic data processing is co-fabricated with the capacitive structure. This co-fabrication is essential for detecting the small capacitance changes associated with near-zero accelerations.

**6.2.4 Tilt sensors (Category 1.5)** or inclination sensors are used in motor vehicles, aircraft, ships and track vehicles. Their purpose is to detect angular displacement with respect to the center of the earth and they are therefore called gravitational sensors by Fraden [27], p.256. Since acceleration is equivalent to changing the direction of gravity, they respond to acceleration as well as tilt. Work done by tilting is similar to the work done on a proof-mass. The mechanisms for implementing tilt sensors are similar to those for accelerometers. The technology may not be the same. An example is the binary tilt sensor used in thermostats, the mercury switch. The non-wetting drop of mercury rolls from one end of the sealed glass capsule to the other end as the capsule is tilted. The electrolytic tilt sensor [27] is very similar in operation, but uses a curved capsule nearly filled with conducting fluid and three electrodes. Its potentiometric voltage divider mechanism provides an analog tilt signal.

**6.2.5 Distance sensors (Category 1.6)**
Distance sensors are distinguished from position (or displacement) sensors because they use a characteristic of a Type M probe such as time-of-flight or wavelength to make the measurement. Probes can be radar, ultrasound and to a lesser extent, optical. The same technologies are used for binary position sensors (Category 1.1). Distance or length is a static characteristic of a structure, and cannot be a source of energy, eliminating the possibility of Type C. Unlike temperature, strain or magnetic field, distance doesn't directly affect material properties, eliminating the possibility of Type P. Calipers **(Category 1.7)** are distance measurement instruments that are also considered sensors. They make use of magnetic resolvers and optical encoders. **Roughness sensors (Category 1.7)** are a small but growing market. Examples are the backscatter of radar probes to measure the roughness of a road sensor and laser backscatter in airplanes to detect clear air turbulence.

**6.2.6 Force (Category 2.10) and Torque (Category 2.11) sensors**
Force transducers are described in Norton [14], chapter 10, which lists various Type C structures for converting a force into a strain or a displacement that can then be measured by a Type M or P transducer. Modern force transducers basically make use of load cells that convert an applied force into the deformation (strain) of an elastic element. The mechanical properties of the load cell are important in determining the accuracy and repeatability of the force measurement. Usually the load cell is made of carefully selected steel. A number of force measurement schemes are possible [14], but the market is dominated (78%) by strain gauge technology based on constantan foil (a copper-nickel alloy with a low temperature coefficient of resistivity). Torque sensors can also be based on a wide variety of mechanisms, but again the market is dominated (70%) by the classic foil-based strain gauge. Brushes make electrical contact to slip rings that connect to the Wheatstone bridge on the rotating shaft. Several brushless approaches, based on the use of magnetic induction pick-off, have been demonstrated [14, 23] and are expected to gain market share. MEMS technology appears unlikely to become a factor in these sensor categories.

**6.3 Mechanical properties related to fluids (Category 2).**
Sensors for mechanical properties related to fluids are quite different than the mechanical properties described above. The measurands are different and the material science around them is very different. They deserve their own category in the sensor marketplace.

### 6.3.1 Pressure (Category 2.1).

A very important measurand in industrial control systems is pressure. Every review of sensors includes a discussion of pressure transducers and several books devote at least a chapter to pressure transducers: Fraden [27], chapter 10, Norton [14], chapter 15 and Soloman [19], chapters 60, 62, 66 and 77. In principle there are many different ways to transform pressure into a strain or displacement. Norton [14] describes nine "pressure-sensing elements." The most familiar are the clamped diaphragm and the C-shaped Bourdon tube that is used in pressure gauges having visual readout. In today's commercial marketplace diaphragms are the only kinds of information transducers used for pressure sensors that have electrical output. Diaphragms can be implemented by silicon micromachining and are a prime example of the success of MEMS [44,45]. We have discussed MEMS pressure sensors that use silicon piezoresistors in a Wheatstone bridge arrangement (See Box 2 above). These dominate the marketplace, but do not meet all application needs. Steel diaphragms with foil strain gauges can operate at higher temperatures than silicon, and dual-stator capacitive transducers (Norton [14], p.305) offer higher over-pressure protection, especially in industrial applications requiring differential pressure measurement. Differential pressure sensors are listed as a separate category (**Category 2.2**) because their markets are different even though their mechanisms are the same. They are largely used to measure liquid level by placing the sensor at the bottom of the container and measuring gauge pressure but are not listed as such in Category 2.3 because it would be double counting.

Although piezoresistive and capacitive transducers dominate the pressure transducer market, resonant strain transducers offer potential advantages. The most interesting is the resonant microbeam strain transducer [12,19,46] that are made by MEMS technology. The transducer elements are clamped-clamped microbeams that can be excited to resonance electrically or optically and whose resonant frequency is highly sensitive to strain. In spite of their attractive features, they have not made a significant impact on the market because the piezoresistive technology is so well developed and has demonstrated high performance at low cost.

Although their primary sensing elements are Type C transducers, pressure transducers do not necessarily have the advantages listed in Box 1. The zero and span stability depend on the material properties that can age or drift with thermal cycling. Particular care must be taken in the isolating the clamped diaphragm from the mechanical packaging materials. Thus these sensors illustrate two points: 1) the generalizations in Box 1 are not absolute and 2) material science is an important aspect of sensor technology.

Methods of sensing pressure that do not use diaphragms are based on entirely different mechanisms. Vacuum gauges measure absolute pressure rather than pressure differences and are limited to pressures well below atmospheric. They are considered instruments, not part of the sensor marketplace. Two kinds are mentioned in Table 4, both Type M. Another kind of pressure measurement is inferred pressure or indirect measurement of pressure. An example is the inference of pressure at the center of the earth using earthquakes as Type M probes. The pressure, density and composition of the earth's core are inferred from models based on geological models and laboratory measurements at high pressures.

### 6.3.2 Fluid flow (Category 2.4)

A variety of measurement principles can be used to measure fluid flow as described, for example by Norton [14], chapter 12 and Allocca [11], section 1.8. Turbine-based flowmeters are Type C flow sensors that convert kinetic energy of the flowing fluid to rotational energy. A familiar example is the turbine on the pumps at gasoline stations where rotating cup anemometers convert fluid flow to mechanical rotation. The rotational motion is then detected magnetically or opto-electronically using Type M transducers. Simple structures such as orifice plates, Venturi meters and pitot tubes are also Type C flow transducers. They convert the flow into a pressure drop that is proportional to the square of the flow velocity. Differential pressure sensors are then used to measure the pressure drop. Again, the markets are counted only once (under differential pressure) in Table 3. These transducers have the advantage of no moving parts, which is desirable for high reliability. A simple "vortex-shedding" structure can be inserted into the flow stream to alternately induce clockwise and counterclockwise vortices. The vortex generation rate depends on the flow velocity and a piezoelectric vibration sensor can detect the frequency of the oscillations.

Flow sensors can also be based on Type M mechanisms. The Coriolis-type flow sensor uses a forced mechanical oscillation as a probe to measure true mass flow. The transverse motion transverse to the flow induces a twist proportional to mass flow that can be measured by an inductive pickoff, giving this sensor a Type M-M classification. The hot-wire anemometer used to measure wind speed in meteorology is a Type M flow sensor that is based on flow-induced modulation of temperature, i.e., a thermal energy probe. The hot wire heats the fluid and the flow carries the heat downstream, giving rise to an increase in power that keeps the heater temperature constant. The mass air flow sensor in the air intake manifold of an automobile engine is based on the same principle, but the hot wire is replace by hot film on a ceramic substrate [18]. Alternatively, the downstream/upstream temperature difference can be detected by a thermocouple (Type C) having the two junctions located symmetrically on the two sides of the heater. A larger signal can be obtained by using a matched pair of temperature-sensitive resistors (Type P) – one upstream and one downstream – in a Wheatstone bridge (Type PM) arrangement. This arrangement is suitable for MEMS implementation in the form of a microbridge [47]. The same device can be used to measure thermal properties of the fluid (namely, thermal conductivity and specific heat) that can be related to properties of interest such as heating value of a gaseous fuel [48].

### 6.4 Thermal quantities (Category 4)

Temperature is without doubt the most commonly input variable sensed and is discussed in every comprehensive review of sensors. See, for example, Fraden [27], chapter 16 and part of 14 and Norton [14], Chapters 19 and 20. Binary temperature sensors such as thermostats are primarily Type C as discussed above in Category 1.0. Thermocouples (also Type C) still represent a large market, even though the technology is very mature. Resistance-based temperature sensors are based on the temperature dependence of electrical resistivity, making them prime examples of Type P transducers. The key to their performance is control of material purity, as exemplified by platinum (-wire) resistance thermometers (RTDs). Semiconductor-based resistance thermometers (thermistors) have many desirable characteristics: small size, high temperature coefficients, fast response time and wide availability at low cost [14]. The market numbers in Table 3 do not do justice to the large volume of devices, because many of them are sold unpackaged at low cost for OEM applications [28]. The performance of the sensor is affected by the package as well as by the degree of thermal contact with the object whose temperature is being measured.

Non-contact temperature measurement is essential for the high temperature ranges (typically up to 2800C), and the art of optical and radiation pyrometry is well developed (Norton[14], chapter 20). Based on development of IR detector technology during the 70's and 80's, the temperature range has been extended to much lower temperatures and the term "radiation thermometry" is appropriate for this increasingly important sensor category. Pyrometers like all other non-contact temperature sensors, require a lens to focus an area of the object being viewed onto the detector, and therefore have Type D primary input transducers as their first element.  MEMS-based fabrication technology has enabled thermal imagers with arrays of detectors and their associated circuitry that can provide "night vision" by detecting the small differences in temperature in a scene that is in total darkness [49].  Silicon-based IR imagers can distinguish differences in temperature less than 0.04C with TV-like resolution and imagers using cooled focal planes can do even better.  Thermal imagers are unique in that the source of information is completely passive.  That is, other imaging sensors require an active energy source in the form of uniform illumination of the scene or a scanning energy probe.

### 6.5 Optical quantities (Category 4).

During the past half-century, the technology of photodetection has completely changed. Silicon technology (reviewed by Middelhoek [3], chapter 2; Norton [14], chapter 22 and Fraden [27], chapter 14) now dominates the scene that used to be the domain of photomultipliers (Type C) and photoconductors (Type P).  Silicon and its oxide not only have excellent properties for microelectronics and integrated circuits, but these properties make silicon ideal for low-leakage photodiodes and photodetector arrays.  Because of its indirect energy gap silicon devices can have good spectral response over the full optical spectrum, a useful property in many applications.  It is the properties of silicon that enable photodetectors to be Type C transducers, having all the attractive features listed in Box 1 above.  Frequently, however they are operated in reverse bias, which puts them in the Type M classification.  Silicon PIN photodiodes have very low leakage currents when operated in reverse bias and thus have most of the advantages of Type C while still providing voltage gain.  Photoconductors (Allocca [11], chapter 12) are Type P devices generally having poorer sensor characteristics, besides being less compact and difficult to integrate into arrays, and are being replaced by silicon photodiodes.

It is the silicon-based imagers (**Category 4.2**), however, that generate most of the data for today's IT systems.  The requirements for high-speed (wide-band) communication systems are driven by the demand for transmitting images, with cameras being even integrated into cell phones.  Linear arrays are widely used for office equipment – scanners, fax machines and copies. The market numbers in Table 3 assume only modest growth in market value, accompanied by a strong deterioration in the price per sensor, because the numbers of sensors sold is growing at a much more rapid rate than the market value. Even with the drop in unit cost the performance (number of pixels, etc.) continues to improve. The market numbers in Table 3 include only sales of sensors sold as such on the open market, they do not include video cameras made as consumer products because these have their own display system included in the product and thus come under the definition of an "instrument."  Neither would it include detectors and arrays made internally by camera or system manufacturers nor the lens unless it is included in the sensor package.  The sensing subsystem mechanism is listed as Type M-D-D-C as discussed in Box 3 with the photodiodes basically operated at zero bias.  The readout is done by applying voltage successively to each element of the array.  This readout out process would be an example of a Type M process but it is performed on the electrical signal and so is not part of the sensing subsystem as we have defined it.  The optical bar-code scanners at checkout counters are listed as Type M-D-M corresponding to transduction processes: the reflection of light, the dispersion by the narrow band filter and the detection by the pin diode.  The laser-scanner is counted as the probe in the Type M process, an auxiliary element that is not in the information flow path. Document readers use linear arrays with cylindrical lenses for one dimension of the image and obtain the other dimension by scanning the light source, lenses and detectors the length of the page.

### 6.6 Acoustic and vibration quantities (Category 5)

In Table 3 we have combined sensors related to vibration, sound and ultrasound. Norton [14], chapter 16 and Fraden [27], chapter 12, discuss the principles of microphones and sound measuring instruments. Allocca [11], chapter 17 discusses vibration sensors and microphones in chapter 17 and devotes chapter 15 to ultrasonic devices. Although vibration sensors and microphones can use a number of different types of transducers, piezoelectric knock sensors for automobile engines dominate the market. These are basically wide-band accelerometers that generate signals for sophisticated data processing [18]. Foil-type electret microphones are also Type C transducers that are high performance sensors, having very wide bandwidth with flat frequency response, low harmonic distortion, low vibration sensitivity, and low noise characteristics ([27], p. 360).

Ultrasonic imagers, like most other imagers, are Type M sensors that use an ultrasonic beam formed by a phased array of transmitters. They make use of the reciprocity property of piezoelectric transducers: the same ceramic elements are used for both generating the ultrasound and receiving the reflected signals. Each sensor has 300-500 piezo-ceramic elements with a trend toward more pixels and higher resolution. A well-known application is for viewing and diagnosing the condition of an unborn child.

### 6.7 Electric and magnetic quantities (Category 6).

A wide variety of mechanisms that can be used for magnetic sensing are described in Ripka's book [23]. However, most of the mechanisms apply to instruments that are not considered part of the sensor market and are not included in Table 3, even though they involve electric and magnetic sensing subsystems. Magnetic sensors are involved in measuring many nonmagnetic measurands: proximity, position, navigational direction, angular speed (rpm), torque and even liquid flow and are listed under those Table 3 categories. Thus Category 6, which refers to sensors sold for the purpose of measuring electrical and magnetic quantities themselves, significantly understates the importance of information transducers based on magnetism. No binary sensors in Category 6 were found, but magnetic mechanisms are prevalent in binary sensors for other measurands.

Electrical and magnetic quantities are closely related through Ampere's law, which gives the magnetic field associated with an electrical current. Current sensing with a shunt resistor is the obvious method of choice, but in some cases it is impractical or impossible. An example is current sensors for high voltage lines. In such cases the magnetic field associated with the current is measured using transformers, Hall effect or magnetoresistive sensors (Norton [14], chapter 24). The conductor carrying current passes through a gapped ferromagnetic toroidal core that concentrates the magnetic field and the Hall device is mounted in the gap. The core is a Type C transducer with electrical energy in the conductor being converted to magnetic field energy in the core.

The market for Category 6 is totally dominated by the read heads used in magnetic information storage systems for the IT market: computers, office equipment and consumer electronics. This technology is quite sophisticated with continuing evolution toward higher and higher data densities [23]. The mechanisms fall into the three categories listed in Table 3 and are briefly described in [9] as well as in books on magnetic storage and magnetic recording [50,51]. Inductive heads are Type M because the moving magnetic media causes a variable magnetic flux that generates the pick-off voltage as discussed above in the section on Type M transducers. Magnetoresistive read heads (Type M-PM), especially those with the giant magnetoresistance (GMR) effect are replacing the inductive read heads because they provide a larger signal and avoid the problem associated with Type P by operating in a fairly narrow bandwidth. Consequently, the market numbers in Table 3 are in a rapid state of flux. Ripka ([23], chapter 4) discusses the GMR effect and its applications to various magnetic sensors and concludes with a highly optimistic statement: "The density and detection rate are increasing at an incredible pace, and GMR read-heads are the best bit sensor for the foreseeable future."

### 6.8 Qualities of materials, liquids and the environment (Category 7)

This category of sensors relates to physical properties since chemical and biological properties of liquids and gases are covered in categories 8 and 9. Many diverse technologies are involved. The main application for these sensors is in the research and development laboratory and other industrial laboratories such as in the process industries, mining and the basic materials sector. These are laboratory instruments and are not included in Table 3 although some are listed in Table 4. Ultrasonic imagers could possibly be in this category since they sense material qualities – density variations that affect acoustic impedance. However, it seems more reasonable to classify them with sound and vibration sensors in Category 5.2. The same argument could be applied to optical imagers, because they sense the optical qualities of objects being imaged, but again the sensor technology being used is that of photodetectors and they are therefore placed in Category 4.

The significant market segments identified in Category 7 are simple low-cost detectors and sensors, all based on Type M mechanisms. Smoke detectors have proven their effectiveness in saving lives and dominate the market in this category, with building codes requiring them in most rooms of most dwellings. Along with rain sensors, which represent a rapidly growing market, they have been discussed above in the section on binary sensors. Rain sensors are difficult to categorize because they are binary (presence/absence of rain activates the windshield wiper), they relate to a liquid quality and an environmental quality and they sense the presence of a chemical, but category 7.0 seems a good choice. Sensors for oil quality in automobiles are based on the increase in electrical conductivity of the oil while the turbidity of washing-machine water is measured by IR scattering [52].

### 6.9 Chemical quantities (Category 8)

Chemical sensors produce an electrical signal in response to the presence of a particular analyte, that is a molecular, ionic or atomic species. In principle, this category covers all analytes for which analytic methods exist. Since such methods include IR and visible spectrometry, gas and liquid chromatography the list of analytes is in the hundreds of thousands, even millions. The chemical sensor faces the formidable task of both identification and quantification of small quantities of analytes that may be present in either the liquid or gaseous matrix. Biological activity is usually associated only with the liquid state, but the complexity of biological molecules presents an even bigger sensing challenge. Nature has managed to meet this challenge much better than current technology.

### 6.9.1 Humidity (Category 8.1)

Humidity is a chemical analyte so unique, pervasive and important that it deserves its own subcategory. It is second only to temperature as an important quantity to measure and control in many industries, buildings and household appliances (clothes dryers, microwave ovens and dessicators). About 20% of this market consists of binary sensors (category 8.0) but the technology is the same whether analog or binary. It is an important aspect of energy management in HVAC systems, of comfort control and of storage and preservation, since many materials are humidity-sensitive. This very humidity sensitivity is the basis for the Type P approaches to humidity sensing listed in Table 3. These and other approaches are described in Soloman [19], chap 39. Humidity sensors are generally low accuracy devices subject to hysteresis, aging and contamination effects. The dew-point sensor or chilled mirror hygrometer is more reliable than the Type P devices and, when combined with temperature cycling of the mirror about the dew-point, provide measurements that can be made traceable to NIST [19]. The temperature cycling means that the measurement principle is a double probe mechanism (Type MM) because temperature is used as a first probe and an optical probe is used to detect condensation/evaporation on the mirror. If the light is detected by a reverse biased photodiode, the classification becomes MM-M.

### 6.9.2 Sensors for ions in liquids (Category 8.2)

The humidity sensors discussed above represent only one analyte. The entire field of chemical sensors covers thousands of analytes and requires mechanisms that can separate and quantify chemical or molecular species in different environments. Many different mechanisms and technologies are used for chemical sensors [53]. The environment determines the appropriateness of a given technique. Sensors for ions in liquids are almost entirely based on electrochemical methods, with the liquid being an aqueous solution (electrolyte). Hundreds of ions are of interest, so this subcategory actually covers more measurands than all the previous categories combined. Furthermore, in the field of analytical electrochemistry there are many methods for measuring a given ionic species.

Potentiometric electrodes dominate the market: pH electrodes, ion-selective electrodes and gas-permeable membrane electrodes are listed in Table 3. These are immersed in the liquid along with a reference electrode and the voltage difference between the two electrodes is measured. Sensing electrodes for dozens of different analytes are available [54-59]. As discussed in books on analytical electrochemistry [60-63] they are based on a selectively permeable membrane that separates two electrolyte phases and through which (ideally) only the desired analyte can pass. The selective membrane, which is critical to the sensor operation, is a Type D mechanism, completely analogous to a narrow band optical filter that transmits a certain wavelength for detection and reflects other wavelengths. For ion-selective electrodes, the membrane is an ionic (as opposed to electronic) conductor and ideally the membrane current is carried only by the analyte ions with no contribution from other ions. The result is that the potential difference across the membrane is by the Nernst equation, i.e., the voltage across the membrane is related logarithmically to the concentration of analyte in the unknown phase. The sensing electrode voltage relative to a reference electrode is measured by a high impedance voltmeter (a "potentiometric" measurement). Traditionally such measurements are laboratory tests made by comparison to a standard solution (having a known concentration of analyte) before and after the unknown. The Nernstian response is very advantageous because it gives useful data over a large concentration range. In practice, of course, the selectivity of the membrane is not infinite and other species can contribute to the output.

The voltage measurement, even though done conceptually at infinite impedance is a Type C conversion of chemical energy to electrical energy, in the same way a dry-cell voltage is a measure of the difference in oxidation/reduction potential of the two electrode materials. On this basis electrochemical sensors using a selective membrane and potentiometric measurement are listed as Type D-C transducers in Category 8.2. An example is the fluoride ($F^-$) ion-selective electrode, which uses single crystal $LaF_3$ as the ion-selective membrane.

The operation of ion-selective electrodes is more complicated than the ideal concept suggests. For example, a large body of knowledge has been developed around the material science of the classic glass pH electrode that has been in use for almost a century [54]. Silicate glasses were developed that exchange hydrogen ions ($H^+$) between the surface of the glass and the solution in which it is immersed. The glass electrode consists of a thin glass bulb that contains a solution of HCl and a silver wire coated with silver chloride. When the bulb is immersed in a solution whose pH is to be measured, the difference in $H^+$ concentration on the two surfaces of the glass membrane generates a voltage across the glass that is the pH sensor output.  However, examination of the conduction in the glass shows that it is not dominated by $H^+$ transport but rather by $Na^+$ ion transport and the contribution of $H^+$ to the current is essentially zero.  In spite of this, the electrode gives an accurate measure of pH because the $H^+$ ions dominate charge transport in the interfacial regions of the glass since these regions are heavily hydrated.  Thus the classification of the glass membrane as a selective filter followed by an energy converter (Type D-C) is appropriate even though the simple concept is flawed.

### 6.9.3 Sensors for dissolved gases in liquids (category 8.3)

Electrodes for sensing gases dissolved in liquids can be either potentiometric or amperometric. The response for a chemical sensor of any dissolved gas can be expressed as the partial pressure of that gas in an inert atmosphere in equilibrium with the analyte liquid.  The partial pressure is a well-known measure of chemical activity [3,27].  Gas-sensing electrodes involve additional levels of selectivity beyond that used in ion-selective electrodes.  These devices are based on a pH electrode that is isolated from the test solution by a gas-permeable polymer membrane.  A small volume of electrolyte is contained between the polymer membrane and the glass membrane.  Depending on the choice of polymer, small gas molecules such as $SO_2$, $NH_3$, and $CO_2$, can permeate from the test solution through the polymer and react reversibly with the sealed electrolyte to produce changes in pH.  Modulation of the pH by the gas means this transduction involves Type M.  Selectivity is determined by the choice of membrane and electrolyte.  Thus these sensors are classified D-M-D-C.  Both membranes act as filters and the chemical equilibrium reaction transforms one chemical signal (gas concentration) to the modulation of another (pH).  Calcium ion electrodes and "water hardness" electrodes are important commercially and are also based on the use of contained liquids that act as ion exchangers [60] and thus also belong to Type D-M-D-C.

The fact that potentiometric measurements require high impedance measurement of small voltages has motivated the development of the ISFET, or ion-sensitive field-effect transistor for the measurement of pH in aqueous solution [59].  These devices are much more compact and easier to work with than the glass electrode, but the required reference electrode limits their attractiveness.  The pH response of the ISFET has also been found to depend on the surface reactions between the electrolyte and the silicon dioxide dielectric and not from the high mobility of hydrogen ions through the oxide [3].  Both bulk electrodes and ISFETs can be coated with selective membranes to make them responsive to different analytes.

Amperometry is a method in which a constant voltage is applied to an electrode and the resulting current is measured.  The quantitative basis for measurement is Faraday's law: each ion that reacts at the cathode corresponds to one electron in the external circuit so the current is proportional to the partial pressure of the dissolved gas.  The classic Clark cell for sensing oxygen listed in Category 8.3 is an example of an amperometric cell that operates at zero voltage. It has 2 electrodes and acts like a battery with an active electrode that is consumed through electrochemical reaction (Type C).  The silver anode and a platinum cathode are enclosed in a sealed membrane containing KCl electrolyte. Oxygen molecules in the unknown solution selectively diffuse through the membrane and the KCl electrolyte and are reduced at the cathode, while silver is oxidized at the anode to complete the circuit. Here the oxygen-permeable membrane acts as a selective filter but has an even more important role: it determines the rate of oxygen diffusion to the cathode. Its function is analogous to that of an orifice plate, which is used to transform flow to differential pressure.  The membrane, depending on its permeability and thickness, transforms one measurand (concentration of oxygen) to another (oxygen flux) and is thus a Type C process.  The energy for the reaction is supplied by the chemical reaction making this a Type C process so that the classic Clark cell is classified as Type C-C.  The disadvantage of the Clark cell is that the silver anode is consumed when current flows, so the sensor has a limited lifetime and is considered a disposable sensor. The three-terminal version of the Clark cell uses an inert platinum working electrode that is not consumed, but uses an enough applied voltage to make the electrode a sink for incoming oxygen and is therefore listed as Type C-M. Applications include aquaculture, environmental monitoring, industrial processes and blood oxygen during surgery.

Potentiometric electrodes and amperometric cells represent only a fraction of the electrochemical methods of analysis available [60,62]. Most of these methods are laboratory methods rather than sensors that can operate reliably on line without operator intervention. However, it there are significant sensor development efforts under way and it is worthwhile to see how some of these methods fit into our classification scheme. Voltammetry requires an electrochemical cell having three electrodes: the working or sensing electrode, a counter electrode and a reference electrode. The sensing current is a measure of an electrochemical reaction at the sensing electrode, and the voltage at which it occurs depends on the potential between the sensing electrode and the electrolyte as measured by the reference electrode. The shape of the current–voltage response provides information about analytes and electrochemical variables. Voltammetry has many variations but they all involve a Type M mechanism since a scanning voltage is used as a probe. In "stripping voltammetry" the analyte is first deposited on the sensing electrode as the result of a voltage-controlled oxidation or reduction reaction. The potential is then scanned in a direction to remove the deposited material. This method allows determination of very low concentrations of trace metal ions because they can be collected and concentrated at a very slow rate (limited by diffusion to the sensing electrode) and then the stripped rapidly, with the amount collected determined by Faraday's law. This is an example of storing chemical information as an intermediate step in chemical sensing.

### 6.9.4 Sensors for chemical quantities in gases (Category 8.4)

Sensors for chemical quantities in gases are based on a wide variety of technologies and mechanisms. In principle, they cover all the gases and vapors for which analytic methods exist. The present discussion is limited to the well-known gas sensors that dominate the market. Electrochemical sensors for gasses in air are similar to those for gases dissolved in water but contain the aqueous electrolyte and the separated electrodes inside a gas permeable membrane. Since any membrane is also permeable to water, these cells must be operated in a specified humidity range. Two types of cells are listed in Table 3: two-electrode cells for $O_2$ in air and three-electrode cells for $NO_2$, NO and CO. Their mechanisms are similar to corresponding sensors for dissolved gasses.

The largest-selling gas sensor on the market is the familiar oxygen sensor used for combustion control in automobile exhaust systems [18,40,58]. It is an electrochemical sensor that based on the electrochemical properties of yttria-stabilized zirconia (YSZ). This material is a high temperature ionic conductor, with oxygen ions as the conducting species. It is known as the "lambda probe" because its function is to control the air-to-fuel ratio around the stochiometric point (the lambda point) in order to maximize fuel efficiency and minimize air pollution. It is a concentration cell, that is, one electrode is in contact with the analyte exhaust gas, and the other is in contact with outside air. The cell can be used in the potentiometric mode (Type C) or the amperometric mode (Type M) which appears to be the dominant mode in the market. Amperometric sensors have their cathodes coated with a porous layer that determines the rate at which oxygen diffuses to the cathode. When a voltage is applied to the cell it pumps oxygen from the cathode to the anode at a rate determined by the porous coating and the concentration of oxygen. The current reaches a saturation value that is proportional to the concentration of oxygen. Here the porous layer doesn't play the role of a selective filter, but transforms one measurand (concentration of oxygen) to another (oxygen flux) giving these devices a Type C-M classification, similar to the Clark cell. In this case an electrode is not consumed, and the lifetime is limited only by degradation of the materials because of the high operating temperatures.

Semiconductor-based gas sensors for combustible gases are solid-state devices composed of sintered metal oxides which detect gas through an increase in electrical conductivity (Type PM) when reducing gases are adsorbed on the sensor's surface. The best-known and most emulated version is the Taguchi gas sensor (TGS), named after the inventor [39]. The sensing material is primarily polycrystalline $SnO_2$, a wide band-gap n-type semiconductor with resistivity dominated by potential barriers at the grain boundaries. The sensor uses an auxiliary heater that heats it several hundred degrees Celsius above room temperature. The heating lowers the resistance to a reasonable value, gives fast response time and minimizes problems with water vapor, a notorious problem for most gas sensors.

Based on its overall operation the TGS is clearly a PM mechanism. However, a more detailed examination of the mechanism of resistance change leads to a more complex mechanism. At the operating temperature oxygen molecules are split into atomic oxygen, which diffuses rapidly along grain boundaries and is adsorbed at negatively charged interface (or surface) states. The negatively charged grain boundaries increase increasing the height of the potential barriers for the conduction electrons. Combustible gas molecules react with the adsorbed oxygen ions, lowering the barrier height. For example: $CO + O^- => CO_2 + e$. Even small amounts of reducing gas can react with the atomic oxygen, reducing the barrier height and the electrical resistance. In terms of mechanism type, the heater and available oxygen on the catalytic substrate produce atomic oxygen as an energy source that is modulated by the CO analyte and is thus a Type M mechanism. The electrochemical reaction changes chemical energy to electrical energy associated with the potential barrier (Type C) and the potential barrier modulates the conductivity (Type P), which in turn requires another Type M mechanism to generate an electrical sensor signal. Thus the detailed mechanism is M-C-P-M, which may be more appropriate from the semiconductor scientist's point of view than the simple P-M classification given in Table 3.

Additions of other metal oxides to the $SnO_2$ semiconducting oxide affect the relative response to different reducing gases, enabling a family of sensors for common combustible gases, toxic gases like carbon monoxide, ammonia and hydrogen sulfide, and organic vapors. Unfortunately, the selectivity is not perfect: all the sensors respond at least weakly to any gas that can react with oxygen. In spite of this there are many popular applications including alarms for toxic and explosive gases, breath alcohol checkers, automatic cooking controls for microwave ovens, air quality and ventilation control systems for both homes and buildings [39].

Other sensors for either combustible or noncombustible gases are based on other semiconducting oxides such as $TiO_2$, $ZnO_2$, and $WO_3$. These all produce changes in resistivity (Type PM) that are strongest for the analyte gas of interest, but also suffer from cross-sensitivity to other gases. Besides the YSZ oxygen sensor described above, other solid electrolytes that can be used for gas sensors are β-alumina, Nasicon and Nafion. They also require high temperatures for operation and can be operated potentiometrically or amperometrically. Stetter [53] reviews recent developments in all these areas. He also summarizes the development efforts in sensor arrays that make use of a number of sensors responding preferentially to a range of analytes.

### 6.10 Biological quantities (category 9)

Biosensors are sensors that detect and identify molecular species of biological significance for IT systems (as opposed to biology-based sensing systems in living organisms). They could be considered a subset of chemical sensors because they use the same mechanisms as other chemical sensors, but they are developing at such a pace that they deserve their own category. Biosensors generally require some sort of bio-element, for example an enzyme or an antibody, in order to achieve specificity. These function only in an aqueous environment.

### 6.10.1 Glucose sensors (category 9.1)

The best known biosensors are glucose sensors, which are by far the most commercially successful biosensor to date with a world market in excess of about $1.5 billion US including both disposable and non-disposable sensors. They are based on electrochemical mechanisms similar to the Clark cell described above and have undergone three generations of development [64,65]. Although there are many variations in detailed mechanisms, they all are in the category of amperometric enzyme biosensors, with the enzyme being glucose oxidase (GOX). The GOX enzyme molecules are immobilized (covalently attached) at the platinum cathode and are normally in their oxidized state, called GOX-FAD. (FAD refers to the flavin site of GOX, where $FAD^+$ is flavin adenine dinucleotide, an important biomolecule in the Krebs cycle that accounts for the oxidation of most carbon compounds [66,67].) In the simplest version, GOX-FAD reacts with incoming glucose molecules and is transformed to its reduced state, GOX-FADH$_2$, which then quickly reacts with dissolved $O_2$ molecules in a second oxidation-reduction reaction, generating hydrogen peroxide, $H_2O_2$, and regenerating the GOX-FAD. In principle, the oxygen can be detected at the platinum cathode and the glucose concentration can be deduced from the $O_2$ that would be present with no glucose. In practice it is better to detect the $H_2O_2$, which can be done by using the same cathode at the appropriate potential.

The glucose sensor mechanisms outlined above provide interesting examples of our classification approach. The first interface with the liquid to be analyzed is a thin polymer membrane that contains and protects the enzyme. Its purpose is not to provide selectivity or to limit and define the flow (as in the Clark cell), so it should not be counted as an ETP in classifying the sensor mechanism. The reaction of glucose with GOX-FAD converts the signal to GOX-FADH$_2$, a Type C process since it converts the chemical energy of glucose. The next reaction is Type M since it uses the dissolved $O_2$ as the energy source. Converting it to $H_2O_2$ and modulating the $O_2$ signal. Thus, if $O_2$ is detected, the mechanism is C-M-M, where the last ETP refers to the amperometric $O_2$ detection. If $H_2O_2$ is detected, the mechanism is C-C-M. From the general properties of Type C reactions, the second method (in which $H_2O_2$ is detected) results in a better baseline, basically because the $H_2O_2$ is a molecule-for-molecule replacement for glucose. The amperometric (Type M) technique is a weakness in either approach, because other compounds besides $H_2O_2$ can be oxidized at the operating voltage. A second generation of glucose sensors has been designed to address this problem [64].

One might ask, what determines the current if not the permeability of the membrane? As in most redox reactions that are monitored by amperometry, the potential is increased until the reaction rate is limited by the diffusion of the reactant to the electrode (as show by the plateau in the I-V characteristic). In the glucose sensor the diffusion of the glucose in the blood sample and the activity of the enzyme limit the transport. These are sensitive to temperature and a temperature sensor is included for temperature compensation.

### 6.10.2 DNA chips (category 9.2)

Besides enzyme-based sensors, another important category of biosensors is those technologies that use arrays of sensors to detect many variables simultaneously for medical and laboratory diagnostics. Examples are electronic noses and "lab-on-a-chip" analytical systems under development for diagnosis of disease, food quality and homeland security as well as wine, perfume and coffee analysis [53]. The most striking and successful example is the array technology developed for testing DNA samples and DNA sequencing [68]. A DNA microarray (or DNA chip) is a piece of glass or plastic on which single-stranded segments of DNA (probes) have been attached in an x-y matrix array. Each element of the array consists of hundreds of identical segments, making each element of the matrix an effective detector. Typically DNA chips are used to detect the presence of messenger RNA (mRNA) that was transcribed from a sample of genetic material to be analyzed. The mRNA is converted to complementary DNA (cDNA) and "amplified" by the rtPCR process, and fluorescent tags are attached to each cDNA molecule. The probe array is then exposed to the sample containing the cDNA and rinsed. A cDNA molecule that contains a sequence complementary to one of the single-stranded probe sequences will stick to it, causing that element of the matrix to fluoresce when examined with UV light. The whole array can be viewed at once with a video camera and digitized, enabling thousands of genetic tests to be conducted in parallel. The intensity of the fluorescence indicates how many copies of the mRNA were present in the sample and thus roughly indicates the activity or expression level of that gene. In fact, two to four samples can be compared in a single image, since they can be tagged with different colors and exposed to the same gene chip.

The information flow involves a sequence of Type M processes: the conversion of mRNA molecules to cDNA, the amplification of the cDNA, the tagging reactions and the binding to the DNA segments on the probe. The chemical energy for each step is provided by reagents that change the form of the signal. The UV-induced fluorescence is clearly Type M. Thus the probable classification of the overall mechanism is M-M-M-M-M, subject to review by a biochemical specialist.

When combined with sophisticated data processing DNA chips offer a very powerful tool for a variety of applications [53]. In the medical research field they are being used for identifying genes that might be associated with a particular disease, for identifying genetic predisposition to disease, for testing the effectiveness of different drugs at the various steps of drug discovery and in identifying hereditary genetic mutations. In the legal and law enforcement field it is used for forensic applications and paternity identification. The field is part of the rapidly expanding field of bioinformatics or computational biology, including the data mining, analysis and applications of the data gathered by the various genome projects.

Most bioanalytical methods are practiced in a laboratory setting and are not true sensors able to provide reliable electrical inputs to control systems. They can be highly automated but they still require human interpretation, intervention and filtering of information. In fact this statement applies to most of the methods of analytical chemistry. Analysis of such methods is beyond the scope of this chapter, but in the following section we list some of them and suggest how the four basic information transduction mechanisms apply, even if the "sensor" label does not.

## 7. OTHER INPUT TRANSDUCERS AND INFORMATION GATHERING SYSTEMS (Table 4)

Many of the devices, instruments and information gathering systems that generate input data for IT systems are not considered sensors. Nevertheless, they include sensing subsystems that gather and transform information and it is appropriate to consider how they relate to the four basic transduction mechanisms. Table 4, which includes examples of input information transducers that are not part of the sensor marketplace, is intended to be suggestive and is not complete by any means. In the following discussion we highlight some of the table entries and illustrate how a particular technology relates to various measurands and input variables. Since Table 4 is basically a continuation of Table 3, the numbering system is simply continued. The classifications of transduction mechanisms in Table 4 are not necessarily complete but are intended to illustrate the minimum number of transduction processes involved in generating an electrical signal.

### 7.1 Time and frequency sensing devices (Category 10)

Time or frequency (the inverse of time) is a critical input for most IT systems and some sensors. Clocks or frequency references are generally not referred to as sensors, with one exception: Norton's book has a section on "Frequency- and time-sensing devices" that summarizes the mechanisms of these devices [14]. They are clearly information transducers, however; they provide input information for IT systems. In principle, any resonant structure (such as a pendulum, a tuning fork or an LC circuit) can act like a clock if energy is supplied to maintain oscillation at the resonant frequency [69]. Simple LC and even RC oscillators provided adequate performance as the internal clocks for the early microprocessors. Today's quartz crystal oscillators provide much better frequency stability with temperature independence, at very low prices. Box 4 above describes a sensor application. Quartz clocks are simple resonant structures driven by an energy source that can be locked in electronically to the resonant frequency of the vibrating piezoelectric crystal. The impedance of the resonator changes very rapidly with frequency in both amplitude and phase in the vicinity of the resonant frequency and acts like a very narrow band electrical filter. This mechanism is consistent with a Type D mechanism in which a broadband electrical signal excites the very high Q mechanical resonance and the mechanical vibration is then converted back to an electrical signal. The electrical probe and the resonant structure are combined in a single device and constitute a Type MD mechanism. Atomic clocks are based on transitions between energy levels that are determined by the electronic structure of the atom. They can operate on several different principles, all of which fall into the Type M-D classification because they all act as highly selective filters that enable electronic oscillators to lock on to the frequency of the atomic resonance.

### 7.2 Touch-based data input transducers (Category 11)

Touch-based transducers for data input are also not considered part of the sensor marketplace, but are described as "tactile sensors" in the literature and in Fraden's book [27], where the technology is reviewed. They are different from the sensors in Table 3 because their resultant output is always displayed as part of a feedback loop involving the person using them. Tactile sensors such as the membrane switches in keyboards respond to small forces that produce relatively large displacements with a snap-action motion that gives direct mechanical ("touch") feedback that the switch has been actuated. These are binary type transducers, not intended for force measurement, but simply to provide a robust pulse when the threshold force is applied. Since mechanical work must be done to produce the displacement, the mechanism is Type C-M.

Computer "mice" have become quite sophisticated devices that sense very small displacements in the x-y axes. The touch-pad "mouse" on a laptop computer responds to a very light touch by using the capacitance of the finger to report on the presence/absence of a finger and its x-y position, and also responds to a firm tap to provide an additional channel (the z input) of information [70]. Each of these modes requires a probe-type mechanism (Type M) for each degree of freedom.

### 7.3 Transducers for reading stored information (Category 12)

Transducers for reading data stored optically or magnetically are listed in this category. Some aspects of this technology have already been discussed in reference to Table 3, but the mechanisms listed there only referred to the devices sold as sensors; in Table 4 they refer to the complete sensing subsystems. Optical read heads for CDs are described by Lesurf, chapter 9, 10 and 11, as a very readable illustration of the application of information theory to sensors [6].

They are classified as Type MM-M in Table 4, with the first M referring to the laser, the second to the rotation of the substrate and the third to the photodetector. Of course a read head can include lenses, a quarter-wave plate and a polarizing beam-splitter as part of the optical path, but three elements are essential to the information flow. The complete assembled system is sold as a unit, but only the photodetector is included in the sensor market in Table 3, Category 4.1.

Magnetic read heads are somewhat simpler than optical read heads, because the source of the read signal is the magnetic field provided by the media, rather than needing a laser probe. As in the optical sensing subsystem, the substrate movement, whether disk or tape, is an essential part of the readout mechanism. Besides scanning the data, it also produces the modulation necessary to sense the magnetic fields stored in the magnetic media. They are classified as Type M-PM in Table 3 for the magnetoresistive devices.

### 7.4 Analytical instruments and tools for gathering chemical information (Category 13)

Various analytical instruments and other tools for gathering chemical and biological information are included in this category. The intent is to illustrate the variety of methods used for laboratory species identification and quantification and to describe the way that information flow relates to the four identified transduction mechanisms at a broad level that does not necessarily include the details. As in the case of reading stored data, auxiliary devices are needed to implement the sensing subsystem. In this case, instead of a moving substrate, the auxiliary element is the sample injector, a device that injects a pulse of multi-component sample into a separator. The instruments (gas chromatograph, liquid chromatograph, electrophoresis, mass spectroscopy) are based on time dispersion – the time that it takes a certain chemical species to arrive at the detector. The dispersion mechanism is similar to a time-of-flight measurement of distance using an energy pulse as a probe, but in these instruments it is a pulse of the analyte sample. The injector is a pulse modulator (Type M) device and the separator is a dispersion device (Type D) that separates the chemical species into a sequence of pulses ("peaks") reaching a detector. In the case of GC (gas chromatography) the detection may involve different mechanisms such as thermal conductivity, photoionization, flame ionization, all of which require are Type M transducers that generate an electrical output. The GC is unique in that it allows sequential analysis: each of the GC peaks can be further separated and analyzed by a second dispersive analyzer. The techniques of GC-GC and GC-MS have become well-developed methods of analysis.

### 7.4 Measurement instruments and systems for gathering information (Category 14)

This wide-ranging but incomplete listing of instruments is included here to illustrate the variety of sensing systems that are in use today. The classification of sensing mechanism is based on a simplified description of their operation, and is intended to be suggestive and self-explanatory. Vacuum gauges are included to illustrate that diaphragms are not the only way to sense pressure. Vibration, however, is only sensed by Type C transducers whether piezoelectric or spring-proof-mass as listed in Table 3, or by movement of a magnet inside a coil as in a seismometer (Table 4). Astronomical telescopes, like thermal imagers, form images from objects that provide the energy for Type C mechanisms. Their primary sensing element is the objective mirror or lens (Type D). Other imaging systems (cameras, ultrasonic imagers, radar, electron microscopes and MRI systems) are Type M requiring an illumination source or a scanned energy beam.

SQUIDs (superconducting quantum interference devices) are included here because they are exquisitely sensitive sensors of magnetic fields and magnetic field gradients. Robert L. Fagaly [35] describes their operation and applications as sensors. They use DC and small RF fields as probes to lock into unique properties of Josephson junctions with an electrical signal generated by an RF pick-off coil. The appropriate classification is Type MMPC, written as a merged symbol because only one structure, the SQUID coil is involved. SQUIDs must be operated at cryogenic temperatures with complex electronic circuitry but are unchallenged in their sensitivity, which can be in the fentoTesla ($10^{-15}$ T) region. Applications include geophysical mapping, nondestructive test and evaluation and biomedical studies of neuromagnetism and magnetocardiography.

## 8. SUMMARY AND CONCLUSIONS

Information transducers provide the means by which information, whether of human origin or gathered from the exterior world enters IT systems. The information flow can be broken down into a sequence of elemental transduction processes (ETPs). Although sensing systems as a whole are difficult to classify systematically, the ETP mechanisms fall into four quite distinct types: energy conversion, energy dispersion, energy modulation and property modulation (C, D, M and P). Usually identification of the appropriate ETP is obvious; sometimes it is subtle and requires careful consideration of the information flow. In some cases (chemical sensors, in particular) the overall mechanism is readily identified but a detailed look at the process sequence indicates a more complex mechanism. We have found that these four transducer types apply to all information entered into IT systems, that a given type may be used for very different measurands and that information about a given measurand may be gathered by different types of ETPs. This implies that the concept of information transduction is fundamental to the sensing process and perhaps even to the nature of physical information itself.

The "sensor" as defined by the sensor marketplace may not include all the necessary elements of its sensing subsystem that gathers and transforms the information entering an IT system. Or a "sensor" may include electronic signal processing elements that are not part of the sensing mechanism. In Table 3 we examine about 100 different kinds of commercially available sensors, estimate the size of their 2003 global market, and classify their physical mechanisms as a sequence of ETP types. In some cases the classification is oversimplified because a given kind of sensor can be represented by a number of different products. We have sought to identify the mechanism that dominates each market. In Table 4 we list information-input transducers, analytical instruments and measurement systems that are not considered sensor products but are based on sensing subsystems. We show that, at least from an overall simplified viewpoint, they too can be classified according to ETP type. The implication is that there is no fundamental distinction between "sensors" and sensor-like input transducers (e.g., the first column of Table 1) and that the meaning of "sensor" might in the future be broadened to include all input devices for IT systems.

The discussion has been confined to technology-based sensors and has not included nature-based information transduction in sensory systems (vision, hearing, smell, taste, touch, temperature, humidity, $CO_2$, etc,) or in the much broader area of biology-based information systems. Even introductory biology textbooks [66,67] stress the importance of information flow in living organisms, both in interactions with the external world and at the intracellular level.

Biological signals consist of molecules and chemical species as well as neurological signals, making them far more complex than IT's electronic signals. Each living cell contains numerous information transduction and information processing systems, most of which are not understood in detail. Information flow and transduction is involved in reading DNA, in harvesting and using energy, in growth and development and detecting external stimuli. It may well be that the four basic mechanisms we have identified for technology-based systems are found to be inadequate for describing the flow of biology-based information. However, it is usually the case that the understanding of a technology-based system precedes the understanding of analogous biological systems.

The field of sensors covers a very broad base of science and technology ranging from basic physics and material science to biochemistry. The sensors marketplace is highly segmented and coves a wide range of manufacturing and engineering technologies. It is hoped that the concept of information transduction will provide depth to this field as a unifying basis for understanding, explaining and teaching about sensing mechanisms. It is also hoped that it will increase recognition that sensors and sensing mechanisms are aspects of information science and are at the heart of the meaning of physical information.

*Added note:* After this manuscript was accepted for publication, we became aware of a paper by information scientist Bob Losee entitled "A Discipline Independent Definition of Information" [71] that strengthens the viewpoints expressed in this chapter. His key insight is that information is associated with the results of processes: "*information is produced by all processes and it is the values of characteristics in the processes' output that are information.*" An implication is that all physical changes result in the production of information. In the case of an input information transducer the "values of characteristics" refers to the output signal of the transduction process. This output is representative of the input – the physical variable being sensed or transduced, and is informative about the input as well as the transducer. The measuring or sensing process makes information available; it is not created by the measurement process. The breakdown of a sensing process into its sequence of elemental processes (ETPs) is an illustration of "hierarchies of processes, linked together to provide a communication channel." Losee's discussion of information is very broad, from the processes of quantum physics to those of information theory to nonphysical levels including mental processes, not necessarily describable. Without presuming its applicability to other fields, we believe his definition is highly applicable and appropriate to the field of sensors.

**Acknowledgements –** One of us (DZ) acknowledges Honeywell for its long-standing support of R&D in the field of sensing and controls. We are deeply indebted to Peter Dierauer (now at Yellowspring Instruments) for identifying a need for sensor classifications as part of Honeywell's strategic planning and for many helpful insights. Many thanks are due to David Burns, Tariq Samad and Ben Hocker, for critical reading of the manuscript and helpful suggestions, and to Phil Springstead for the artwork. We also acknowledge helpful discussions with Ulrich Bonne, Cleo Cabuz and Don Foreman, all at Honeywell, and Bob Losee at the University of North Carolina-Chapel Hill.

## GLOSSARY

**Control** A device or system that regulates or directs some operation. Usually autonomous except for control set-points determined by human input.
**Dispersion** A process of separating into identifiable components
**Elemental transduction process** (ETP, as defined in this chapter) The smallest possible step in a transduction process.

**Energy**  A quantity that describes the ability to do work.

**Energy conversion**  A transformation of one form of energy to another.

**Information**  Data or values that characterize the output of a process and that are informative about the input to the process as well as the process itself.

**Information processing**  The acquisition, recording, manipulation, organization, storing, retrieval, display, and dissemination of information.  Includes signal processing.

**Information technology**  Technology used for the gathering, processing and distribution of information, including the use of computers and telecommunications.

**Information transduction**  The mapping or transformation of information from one form to another.  Usually the form of energy that carries the information is transformed.

**Modulation**  A process in which a characteristic of one form of energy is varied in accordance with the characteristic of another form of energy.  (See Box 1.)

**Process**  A series (or sequence) of operations, activities or changes organized and conducted to achieve an end result (or output).

**Sensing subsystem** (As defined in this paper)  The set of elements, structures or devices that are involved in gathering information about the physical world and transform it to an electrical signal.  (See Figure 1.)

**Sensing system**  A system that through a series of processes gathers information about the physical world and makes it available for use.

**Sensor**  A device that generates an electrical signal representative of a characteristic or attribute of  the physical world.

**Signal processing**  The transformation, modification or manipulation of a data stream.

**List of Tables and Table Captions**
Table 1. The three parts of a (broadly defined) IT system
Table 2. Measurand categories (table of contents for Table 3)
Table 3. Commercial sensor technologies by measurand category
Table 4. Examples of other input transducers for IT systems (Not considered part of sensor marketplace; numbering continued from Table 3)

**List of Figures** (inserted into text with Figure captions)
Figure 1. Sensor element vs. smart sensors
Figures for Box 1 - symbols for the 4 types and combinations.
Figure for Box 4 – odometer info flow
Figure for Box 3 – camera info flow
Figure for Box 2 – Pressure sensor info flow

**List of boxes**
Box 1. Elemental transduction processes (ETPs).
Box 4. Bicycle odometer
Box 3. Color video camera
Box 2. Smart pressure transducer

## Tables

### Table 1. The three parts of a (broadly defined) IT system

| Input Transducers | Information Processors | Output Transducers |
|---|---|---|
| Touch-based: | Computers | Displays: CRTs, LCDs, etc |
| keyboards | Amplifiers, filters, limiters | Printers |
| computer mice | Mixers | Write heads for storing data |
| Read heads for stored data | Controllers (PID, etc.) | Loudspeakers, headphones |
| Cameras | Counters | Control valves |
| Scanners | Multiplexers | Pneumatic controls |
| Data instruments | A/D converters | Hydraulic controls |
| Sensors | Fourier transforms (FFTs) | Motors, pumps, fans |

### Table 2. Measurand categories (table of contents for Table 3)

| 1. Mechanical quantities related to solid bodies | | | | |
|---|---|---|---|---|
| 1.0 Binary Mechanical | 1.1 Position | 1.2 Navigation | 1.3 Speed and RPM | 1.4 Acceleration/ vibration |
| | 1.5 Tilt | 1.6 Distance | 1.7 Caliper dimension | 1.8 Roughness |
| | 1.9 Thickness | 1.10 Force | 1.11 Torque | |
| 2. Mechanical quantities related to fluids (gases, liquids and powders) | | | | |
| 2.0 Binary fluid mechanical | 2.1 Pressure | 2.2 Differential pressure | 2.3 Level | 2.4 Flow |
| 3. Thermal quantities | | | | |
| 3.0 Binary thermal | 3.1 Temperature | 3.2 Infrared images | | |
| 4. Optical quantities | | | | |
| 4.0 Binary optical | 4.1 Light Intensity | 4.2 Optical images | 4.3 Other radiation | |
| 5. Acoustic and vibration quantities | | | | |

| 5.0 Binary acoustic | 5.1 Sound | 5.2 Ultrasonic images | | |
|---|---|---|---|---|
| 6. Electric and magnetic quantities | | | | |
| 6.0 Binary electric/ magnetic | 6.1 Electric current | 6.2 Magnetic fields | | |
| 7. Qualities of materials, liquids and the environment | | | | |
| 7.0 Binary quality | 7.1 Material quality | 7.3 Liquid quality | 7.4 Environ-mental quality | |
| 8. Chemical quantities | | | | |
| 8.0 Binary chemical | 8.1 Humidity sensors | 8.2 Ions in liquids | 8.3 Gasses in liquids | 8.4 Chemicals in gases |
| 9. Biological quantities | | | | |
| 9.0 Binary biological | 9.1 Glucose sensors | 9.2 DNA chips | 9.3 Other biosensors | |

**Table 3. Commercial sensor technologies by measurand category**

| Category | Measurand or input variable | Sensor or transducer technology | Sensing subsystem mechanism | Comments, Applications, Familiar examples | 2003 Global Market (M $) |
|---|---|---|---|---|---|
| **1. Mechanical quantities not related to fluids** | | | | | |
| 1.0 | Binary position | Limit switches electro-mechanical | **C-M.** Work on spring to actuate on/off switch | Manufacturing controls, automotive | ~500 |
| 1.0 | Binary presence | Light barriers (optical, IR, UV) | **M-M.** Optical beam probe – pin diode | Office automation, Automatic door openers | 1480 |
| 1.0 | Binary presence, position | Magnetic-based proximity switches | **C.** Magnetic induction loops | Traffic control (presence of vehicle) | 80 |
| | | | **M.** AC magnetic field probe | Automotive doors, windows; seatbelts, | 600 |
| | | | **C-M.** Hall-based**.** | | 280 |
| 1.0 | Binary position | Ultrasonic proximity switches | **M-C.** Ultrasonic probe – piezoelectric | Intrusion alarm, also used for analog distance | 130 |
| 1.0 | Motion | Change in reflected light | **M-M.** IR probe – pin photodiode | Security, energy savings | 1420 |
| 1.0 | Airbag sensors (crash event) | Piezoelectric | **C-C.** Work done on proof mass - piezo | Automotive crash sensors –airbags, etc | 130 |
| | | Capacitive | **C-M.** | " | 180 |
| | | Piezoresistive | **C-PM** | " | 50 |
| 1.1 | Linear position | Encoders (optical) | **M-M.** optical probe – pin photodiode | Industrial robots, machine tools | 1500 |
| 1.1 | Angular position | Resolvers (magnetic) | **M-C**. AC magnetic probe - coil | Manufacturing, process control | 950 |
| 1.1 | Position | Potentiometer (variable voltage divider) | **C-M.** Work done on wiper – voltage probe | Automobile gas pedal, throttle valve | 900 |
| 1.1 | Position | LVDT – linear variable differential transformer | **M-C.** AC magnetic probe - coil | Automotive, aerospace, manufacturing | 100 |
| 1.1 | Position | Variable reluctance | **M-C.** AC magnetic probe - coil | Angular position | 150 |
| 1.2 | Position (navigation) | GPS receiver | **M-C.** EM pulses from GPS satellites - receiver antenna | Measures position with respect to satellites | 400 |
| 1.2 | Direction (navigation) | Ring laser gyros | **M-M**. Laser interference – photodiode | Inertial reference systems for aerospace | 850 |
| 1.2 | Yaw | Silicon MEMS | **M-M.** Oscillatory | Inertial reference | 650 |
| | (navigation) | gyro | rotation probe – capacitance pickoff | systems | |
| 1.2 | Direction (navigation) | Fiber-optic gyros | **M-M.** Laser interference – photodiode | Inertial reference systems for aerospace | 100 |
| 1.2 | Direction (navigation) | Mechanical gyros | **M-M.** Angular momentum - capacitance | Inertial reference systems | 100 |
| 1.2 | Direction (navigation) | Magnetic compass | **PM.** Magneto-resistance | Automotive compass | 80 |
| 1.3 | Speed & RPM | Tachometer | **C.** Mechanical to electrical (magnetic induction) | Frequency and amplitude proportional to RPM | 340 |
| 1.3 | Speed & RPM | Variable reluctance | **M.** AC impedance | Same sensor gives angular position | 550 |
| 1.3 | Speed and RPM | Optical encoders | **M-M.** Optical probe - photodiode | Based on counting pulses | 170 |
| 1.3 | Speed & RPM | Hall (based on pulse counting) | **M-M.** Magnetic field - Hall effect | Automotive ABS, ignition timing | 200 |
| 1.4 | Acceleration/ vibration | Piezoelectric | **C-C.** Work done on proof mass | Non-automotive applications | 240 |
| | | Capacitive | **C-M** | | 180 |
| | | Piezoresistive | **C-PM** | | 150 |
| 1.5 | Tilt | Magneto-inductive tilt sensors | **C-M.** Work on mass - Magnetic field probe | Also responds to acceleration | 50 |
| 1.5 | Tilt | Potentiometric tilt sensors | **C-M** Voltage divider | Bubble in a liquid electrolyte | 46 |
| 1.5 | Tilt | Tilt sensors based on accelerometer | **C-C, C-M, or C-PM** | Accelerometer technology (see above) | 70 |
| 1.6 | Distance | Pulsed radar | **M-C.** Microwave time of flight probe - antenna | Time of flight. Also gives velocity by Doppler effect | 750 |
| 1.7 | Caliper distance | Optical calipers (encoders) | **M-M.** Optical probe - photodiode | Optical scale/grating, | 12 |
| 1.7 | Caliper distance | Inductive calipers (resolvers) | **M.** Inductive impedance | Inductive LVDT | 14 |
| 1.8 | Roughness | Roughness | **M-C.** Microwave probe - antenna | Road roughness – back-scatter | 6 |
| 1.9 | Thickness | Thickness | **M-C.** Ultrasonic probe | Time of flight, interferometry | 60 |
| 1.9 | Thickness | Thickness | **M-C.** Gamma radiation probe | Back-scattered radiation | 16 |
| 1.10 | Force (weight) | Foil strain gauge | **C-PM** Work done against gravity – electrical resistivity | Load cell is analogous to diaphragm in pressure transducer | 420 |
| 1.11 | Torque | Foil strain gauge | **C-PM** Work done | Aircraft, shipbuilding, | 110 |

| | | | on torsion bar – electrical resistivity | rail, automotive. Brush contacts to shaft. | |
|---|---|---|---|---|---|
| 1.11 | Torque | Magneto-elastic effect | **C-PM** Torsion bar - Strain - change in magnetic properties | Brushless operation | 20 |
| | | | | | |
| **2. Mechanical quantities related to fluids: gases, liquids and particulates** | | | | | |
| 2.0 | Binary level | Float Reed switch | **C-M.** Fluid does work to activate switch | Level monitor | 100 |
| 2.0 | Binary liquid flow | Flow monitor | Various mechanism(s) | Analog output converted to binary | 130 |
| 2.1 | Pressure | Capacitive | **C-M.** Work on diaphragm – AC probe | Built-in over-pressure protection | 620 |
| 2.1 | Pressure | Silicon piezoresistive bridge | **C-PM.** Work on diaphragm – DC resistive probe | MEMS-based (See Box 4) | 1280 |
| 2.1 | Pressure | Foil strain gauge | **C-PM.** Work on diaphragm – DC resistive probe | Constantan (copper-nickel alloy) foil | 650 |
| 2.2 | Differential pressure | Capacitive | Same as pressure above | | 630 |
| 2.2 | Differential pressure | Silicon piezoresistive bridge | Same as pressure above | | 390 |
| 2.2 | Differential pressure | Strain gauge | Same as pressure above | | 230 |
| 2.3 | Level | Float potentiometric | **C-M** Liquid does work on the float. | Level measurement. | 280 |
| 2.3 | Level | Ultrasonic, radar, laser | Ultrasonic, radar, laser beam probes | Same mechanisms as distance sensors | 320 |
| 2.3 | Level | Fiber optic in liquid | **M-M.** Optical probe - photodiode | Also used for temperature sensing | 10 |
| 2.3 | Level | Capacitive | **M.** Partially filled capacitor | Dielectric fliuds | 300 |
| 2.4 | Liquid flow | Orifice plate | **C.** Flow converted to differential pressure | Market already included under differential pressure | NA |
| 2.4 | Flow (Gas & Liquid) | Rotameter (turbine) | **C-M.** Flow to rotation - magnetic | | 100 |
| 2.4 | Flow (Gas & Liquid) | Vortex | **C-C.** DC Flow to oscillatory flow – piezoelectric | Sense frequency of oscillations | 190 |
| 2.4 | Liquid mass flow | Coriolis Force | **M-M.** Driven oscillation – inductive pickoff | Measures true mass flow. | 300 |
| 2.4 | Liquid flow | Magneto inductive flow | **M.** Magnetic field probe (generates | Only for conducting liquids – water, acids, | 260 |

| | | | transverse voltage) | etc. | |
|---|---|---|---|---|---|
| 2.4 | Gas mass flow | Anemometer (hot wire, hot film and MEMS microbridge) | **M-PM**. Heat probe – RTD pickoff | Microbridge can also be used for gas composition sensing [48] | 1000 |
| 2.4 | Particulate flow (Bulk Material) | Load cells | **C-MP.** Electrical resistivity | Container filling | ~200 |
| | | | | | |
| **3. Thermal quantities** | | | | | |
| 3.0 | Binary temperature | Bimetal thermostats | **C-M** Thermal to mechanical to electrical | Binary output by mechanical switching | ~3000 |
| 3.0 | Binary temperature | Capillary temperature switch | **C-M** Thermal to mechanical - electrical | Temperature monitor and control | 500 |
| 3.0 | Binary temperature | Temperature monitors | **PM.** Thermistors, RTDs | Analog output converted to binary electronically | 160 |
| 3.1 | Temperature difference | Thermocouple | **C.** Peltier effect generates voltage | Process monitoring and control | 1000 |
| 3.1 | Temperature | RTDs | **PM.** Electrical resistivity | Process monitoring and control | 1100 |
| 3.1 | Temperature | Thermistor | **PM.** Electrical resistivity | Process monitoring and control | 1330 |
| 3.1 | Non contact temperature | IR silicon | D-C Lens - Thermopile or **D-PM.** Thermistor | Silicon/MEMS-based | 210 |
| 3.1 | Non contact temperature | IR pyrometer | **D-PM.** Lens - bolometer | Measures temperature rise at focal plane | 110 |
| 3.2 | IR imaging | Silicon microbolometer arrays | **D-PM.** Lens - Electrical resistivity | Used for heat energy measurements, night vision | 130 |
| | | | | | |
| **4. Optical quantities** | | | | | |
| 4.0 | Binary light intensity | Photoconductors | **PM.** Resistivity modulation | Outdoor, building lighting controls | 120 |
| 4.1 | Light Intensity | Silicon PIN & phototransistors | **M.** Reverse bias photodiode | Stored optical data, CD, DVD, CD-ROM | 830 |
| 4.1 | Light intensity | Photomultiplier | **M.** Photoemission | Instruments, chemical analysis | 110 |
| 4.2 | Optical images | Silicon CCD or CMOS 2D arrays | **M-M-D-C.** (See Box 3) | Consumer digital cameras, security systems, microscopes, telescopes | 1400 |
| 4.2 | Optical bar-code | Bar code scanners | **M-D-M**. Scanned laser – filter - pin diode | Product identification, checkout counters. Market includes assembled system. | 1600 |
| 4.2 | Optical image (linear arrays) | Document scanners | **M, D-C**. Scanner, lens - photodiode arrays | Office copiers, image digitization. Market includes diodes only | 380 |
| | | | | | |
| **5. Acoustic and vibration quantities** | | | | | |
| 5.0 | Event – binary acoustic | Glass breakage sensors | **C**. Mechanical to electrical (piezo-electric, etc) | Security alarm | 140 |
| 5.1 | Mechanical vibration in solids | Piezoelectric vibration sensors | **C**. Mechanical to electrical | Automotive knock sensors, Vibration control and measurement. | 350 |
| 5.1 | Sound (acoustic waves in air) | Piezoelectric and electret microphones | **C**. Mechanical to electrical | Microphones | 235 |
| 5.2 | Medical diagnosis | Ultrasonic imaging | **M-C**. Ultrasonic beams used as probes - detectors | Piezoelectric arrays for density/structural information | 950 |
| | | | | | |
| **6. Electrical and magnetic quantities** | | | | | |
| 6.1 | Electrical current | Catheter electrodes | **C**. Current to voltage | Medical measurements | 72 |
| 6.1 | Electrical current | Shunts for high currents | **C**. Current to voltage | Power measurement and control | 67 |
| 6.1 | Electrical current | Hall- based current sensors | **C-M**. Current to B field - Hall effect | Measurement and control | 90 |
| 6.2 | Data stored on magnetic media | Magneto-inductive | **M-C**. Changing flux induces electrical voltage | Computers and office equipment | 230 |
| 6.2 | Data stored on magnetic media | Magneto-resistive | **M-PM** Electrical resistivity | Computers and office equipment | 180 |
| 6.2 | Data stored on magnetic media | Giant magneto-resistive (GMR) | **M-PM** Electrical resistivity | Computers and office equipment | 820 |
| | | | | | |
| **7.  Qualities of materials, liquids, the environment** | | | | | |
| 7.0 | Event - presence of particulates | Fire (smoke particles) | **M-C**. Optical or radioactive probe | Safety alarm – binary event | 870 |
| 7.0 | Rain on windshield | Rain sensor | **M-C**. Infrared probe | Windshield wiper control | 55 |
| 7.0 | Metal particles in oil | Oil quality | **M**. Electrical conductivity | Automobile maintenance | 50 |

| 7.0 | Water quality | Turbidity | **M-M**. Optical probe - pin diode | Washing machines | ~50 |
|---|---|---|---|---|---|
| | | | | | |

**8. Chemical quantities in liquids and gases**

| | | | | | |
|---|---|---|---|---|---|
| 8.0 | **Binary humidity sensors** | Capacitive or resistive | **PM** Dielectric or resistance change | Comfort, process control | 120 |
| 8.1 | Humidity | Humidity-Capacitive | **PM** Dielectric and Dimensional change | Comfort, process control | 320 |
| 8.1 | Humidity | Humidity-Resistive | **PM** Electrical resistivity | Comfort, process control | 130 |
| 8.1 | Humidity | Dew-point sensor | **MM-M** Temperature modulation - optical probe - photodiode | Process control | 30 |
| 8.2 | Ions in aqueous solution | Ion Selective Electrodes | **D-C**. Selective membrane / potentiometric | Quality control, process control | 780 |
| 8.2 | pH in aqueous solutions | pH electrode | **D-C**. Selective membrane / potentiometric | Quality control, process control | 390 |
| 8.3 | Dissolved gases | Gas-selective electrodes | **D-M-D-C.** Selective membranes / potentiometric | Quality control, process control, water quality | 650 |
| 8.3 | $O_2$ in aqueous solution | Clark cell | **C-C or C-M** Permeable membrane - amperometric | 2-terminal or 3-terminal | 150 |
| 8.4 | NO, $NO_2$, CO, $CO_2$ in air | Electrochemical - other gases | **D-M-D-C** or **C-M.** selective membrane-amperometric | Quality control, process control | 300 |
| 8.4 | $O_2$ in air | Electrochemical Oxygen (Lambda probe) | **C-M.** Chemical to electrical (assuming amperometric) | Automotive exhaust, combustion control, | 1100 |
| 8.4 | Combustible gases | Tin oxide –based (Taguchi gas sensor) | **PM.** Electrical resistivity | Safety. | 210 |
| 8.4 | Gas composition | Semiconductor-noncombustible | **PM.** Electrical resistivity | Quality control, process control | 110 |
| | | | | | |

**9. Biosensors**

| | | | | | |
|---|---|---|---|---|---|
| 9.1 | Blood glucose | Glucose biosensor | **C-C-M.** Chemical reaction-amperometric | Control of diabetes (Disposable and nondisposable) | 1500 |
| 9.2 | Medical | Antibody/DNA | **M-M-M-M-M.** | Identify chemical or | 800 |

| | diagnostics | (Gene chips) | Chemical reactions fluorescent tags | biological information (See Section 6.10.2) | |
|---|---|---|---|---|---|

**Table 4. Examples of other input transducers for IT systems**
**(Not considered part of sensor marketplace; numbering continued from Table 3)**

| Cate gory | Input variable of interest | Transducer technology | Transduction mechanism(s) | Comments, Applications, familiar examples |
|---|---|---|---|---|
| | | | | |
| **10. Time and frequency sensing devices.** | | | | |
| | Time interval | Quartz clocks | **MD.** High-Q resonant piezoelectric filter | Wrist watches, frequency references |
| | Time interval | Atomic clocks | **M-D.** (overall) Microwave energy | Frequency standard |
| | | | | |
| **11. Touch-based data input transducers (tactile sensors)** | | | | |
| | Touch - binary | Keyboards | **C-M.** Electrical switch | Array of binary switches for data entry |
| | Displacement/ movement | Computer mouse | **C-M, C-M.** (x, y inputs) electrical or optical | Motion sensors for data entry |
| | Touch/ movement | Computer touchpads | **M, M, C-M.** (x, y, and z inputs) | Touch-based position and data entry |
| | | | | |
| **12. Transducers for reading stored information.** | | | | |
| | Data stored optically | Optical read heads | **M-M-M.** Laser probe with substrate rotation – pin diode | CD, DVD, CD-ROM |
| | Data stored magnetically | Magnetic read heads | **M-PM** Magnetoresistance | Part of sensor market (See Table 3) |
| | Product Identification | RFID | **M.** RF probe | Inventory control, security |
| | Product Identification | Optical barcode readers | **M-D-M** Scanned laser- pin diode | Part of sensor market (See Table 3) |
| | | | | |
| **13. Analytical instruments and tools for gathering chemical information** | | | | |
| | Molecular and atomic spectra | Optical, IR and UV spectrometry | **M-D-M.** Optical probe - dispersive element - detector | Analytical chemistry, species identification and quantification |
| | Chemical species in gas or liquid matrices | Gas and liquid chromatographs | **M-D-M** Valve injector - dispersive column – detector. | Separates molecules according to their adsorption-desorption behavior |
| | Chemical or biological species | Gel electrophoresis | **M-D-M** Valve injector - dispersive column – detector | Separate biological molecules according to their mobility |
| | Atoms or molecular | Mass spectrometers | **M-M-D-M.** Injector – ionizer – | Separates molecular fragments according |

| | fragments | | magnetic dispersion – detector | to their mass to charge ratio |
|---|---|---|---|---|
| | Chemical spectroscopy | Magnetic resonance (ESR and NMR) | **M-M**. RF and DC magnetic field probes | Identify chemical or biological information |
| | | | | |

**14. Measurement instruments and systems for gathering information**

| | | | | |
|---|---|---|---|---|
| | Pressure | Vacuum gauges | **M** Thermal or electrical conductivity probe | Thermocouple gage or ion gage. |
| | Vibration | Seismometers | **C**. Mechanical to electrical | Earthquake detection, oil exploration |
| | Material quality | Viscosity | **M**. Forced rotation as probe | **Quality control** |
| | Color | Color sensors | **M-D-M.** Optical probe- dispersion- detection | Quality control |
| | Magnetic field | Flux gate magnetometer | **M-PM**. AC magnetic probes | Geomagnetic surveys |
| | Magnetic flux | SQUIDs | **MMPC**. DC & RF probes - Josephson effect – RF coil pickoff | R&D tool, |
| | Medical imaging, diagnosis | X-ray machines and CT scanners | **M-C**. X-ray used as probe of structure – X-ray detector | 2D and 3D structural information |
| | Medical imaging | MRI | **M-M.** DC magnetic and RF field probes | Medical diagnosis |
| | Surface structure, composition | Electron microscopes | **M**. Electron beam probes | R&D tool |
| | Surface texture, composition | STM, AFM | **M-M**. Scanning microprobe – optical or capacitive detection | R&D tool |
| | Radioactivity | Radiation, particles | **C**. Particle energy to electrical or optical (pair production) | Instruments for safety |
| | Radioactivity | Radiation, dose | **PM.** Color center formation | Dosimeters |
| | Astronomical images | Telescopes | **D-C.** Mirror – detector array | Photon counting |
| | Astronomical spectra | | **D-D-C.** Mirror - dispersion- photomultiplier | Photon counting |